\documentclass[12pt]{article}
\usepackage{amsmath,amssymb}
\usepackage[english]{babel}
\usepackage[cp866]{inputenc}
\usepackage{hyperref}

\numberwithin{equation}{section}

\newcommand{\bq}{\begin{eqnarray}}
\newcommand{\eq}{\end{eqnarray}}

\newcommand{\bbq}{\begin{equation*}}
\newcommand{\eeq}{\end{equation*}}

\newcommand{\ra}{\rightarrow}
\newcommand{\w}{{\cal W}}

\newcommand{\ov}{\overline}
\newcommand{\nd}{{\ov N}_c}
\newcommand{\oym}{{\ov \Lambda}_{YM}}
\newcommand{\la}{ \Lambda_Q }
\newcommand{\ld}{ \Lambda_q }

\newcommand{\cm}{{\cal M}_{\rm ch}}
\newcommand{\cma}{{\cal M}^{\it l}_{\rm ch}}

\newcommand{\bo}{\rm{b_o}}
\newcommand{\bd}{\rm {\ov b}_o}
\newcommand{\lym}{\Lambda_{YM}}

\newcommand{\ql}{Q_{\it l}}
\newcommand{\oql}{{\ov Q}^{\it \ov l}}
\newcommand{\qh}{Q_{\it h}}
\newcommand{\oqh}{{\ov Q}^{\it \ov h}}
\newcommand{\dl}{{q}^{\it l}}
\newcommand{\odl}{{\ov q}_{\it \ov l}}
\newcommand{\hd}{q^{\it h}}
\newcommand{\ohd}{{\ov q}_{\it h}}
\newcommand{\mh}{m_{\it h}}
\newcommand{\ml}{m_{\it l}}
\newcommand{\hml}{\hat{m}_{\it l}}

\newcommand{\mgl}{\mu_{\it gl,\,l}}
\newcommand{\mgh}{\mu_{\it gl,\,h}}
\newcommand{\mhp}{m^{\rm pole}_{\it h}}
\newcommand{\mlp}{m^{\rm pole}_{\it l}}
\newcommand{\nl}{ N_{\it l}}
\newcommand{\nh}{ N_{\it h}}
\newcommand{\mulp}{\mu^{\rm pole}_{q,\,\it l}}
\newcommand{\hlp}{{\hat\mu}^{\rm pole}_{q,\,\it l}}
\newcommand{\olp}{{\ov\mu}^{\rm pole}_{q,\,\it l}}
\newcommand{\hhp}{{\hat\mu}^{\rm pole}_{q,\,\it h}}
\newcommand{\muhp}{\mu^{\rm pole}_{q,\,\it h}}

\newcommand{\mq}{ m^{\rm pole}_Q}
\newcommand{\ogh}{{\ov \mu}_{gl,\,\it h}}

\newcommand{\mg}{\mu_{\rm gl}}
\newcommand{\sq}{\textsf{Q}_{\it h}}
\newcommand{\oq}{\ov{\textsf{Q}}_{\it h}}

\begin{document}

\begin{center}
{\bf  Mass Spectrum in SQCD and Problems with the Seiberg Duality\,. \\ \, Another Scenario.}
\end{center}
\vspace{1cm}
\begin{center}\bf Victor L. Chernyak $^{a,\,b}$ \end{center}
\begin{center} a) Novosibirsk State University, 630090 Novosibirsk, Russia \end{center}
\begin{center} b) Budker Institute of Nuclear Physics, 630090 Novosibirsk, Russia \end{center}
\begin{center}(e-mail: v.l.chernyak@inp.nsk.su) \end{center}
\vspace{1cm}
\begin{center}{\bf Abstract} \end{center}
\vspace{1cm}

The ${\cal N}=1$ SQCD with $SU(N_c)$ colors and $N_F$ flavors of light quarks is considered
within the dynamical scenario which assumes that quarks can be in the two different phases
only\,:  the HQ (heavy quark) phase where they are confined,\,  or
they are higgsed, at the appropriate values of the Lagrangian parameters.

The mass spectra of this (direct) theory and its Seiberg dual are obtained and compared for
quarks of small equal or unequal masses. It is shown that in those regions of the parameter space where an
additional small parameter exists (it is $0<(3N_c-N_F)/N_F\ll 1$ at the right end of the
conformal window where the direct theory is weakly coupled in the vicinity of its IR-fixed point, or
its dual analog $0<(2N_F-3N_c)/N_F\ll 1$ for the dual theory at the left end of the conformal window), the mass spectra of the direct and dual theories are parametrically different. A number of
other regimes are also considered.

\tableofcontents
\numberwithin{equation}{section}

\newpage

\section{ Introduction}

{\hspace {0.5 cm}} The dynamics of 4-dimensional strongly coupled non-Abelian gauge theories
is complicated. It is well known that supersymmetry (SUSY) leads to some simplifications
in comparison with the ordinary (i.e. non-SUSY) theories. Besides, it is widely believed that
SUSY is relevant to the real world. In any case, it is of great interest to study the
dynamics of the nearest SUSY-relative of the ordinary QCD, i.e. the ${\cal N}=1$ SQCD.
But even with ${\cal N}=1$ SQCD, there is currently no proven
physical picture of even main nontrivial features of its dynamics.

The best proposal so far seems to be Seiberg's dual theory \cite{S1}, which is weakly coupled
when the direct one is strongly coupled, and {\it vice versa} (see
e.g. reviews \cite{AKMRSV, SV-r} for ${\cal N}=1$ SQCD and  \cite{IS, Shif-rev}
for Seiberg's dual theory). The Seiberg duality passed a number non-trivial checks (mainly, the 't Hooft
triangles and the behavior in the conformal regime) but up to now, unfortunately, no proof
has been given that the direct and dual theories are (or are not) equivalent. The reason
is that such a proof needs real understanding of and the control over the dynamics of both theories.
Therefore, in the framework of gauge theories, the Seiberg proposal remains a hypothesis up to now.
We quote here three papers only.  1)\, \cite{AR} (with N. Seiberg among the authors):
"By now many examples of such dualities have been found, and a lot of evidence has been collected
for their validity. However, there is still no general understanding of the origin of these
dualities, nor a prescription to find the dual for a given gauge theory". 2)\, \cite{G}:\, "The most established example of such a duality is the Seiberg duality in ${\cal N}= 1$ SQCD...
The validity of this duality is still a conjecture but it have anyway remarkably passed numerous checks...". 3)\, \cite{A}:\, "All of these dualities do not have any rigorous derivation so far, but they pass many
consistency checks...".

As for the string theory, we quote here two papers. 4)\, The review \cite{K}:\, "Specifically, we have seen using branes that the quantum moduli spaces of vacua and quantum chiral rings of the electric and magnetic SQCD theories coincide. This leaves open the question whether Seiberg's duality extends to an equivalence of the full infrared theory, since in general the chiral ring does not fully specify the infrared conformal field theory. It is believed that in gauge theory the answer is yes, and to prove it in brane theory will require an understanding of the smoothness of the transition when fivebranes cross. It is important to emphasize that the question cannot be addressed using any currently available tools". 2)\, The most detailed discussion of implicit dynamical assumptions and weak points in attempts to derive Seiberg's duality with a help of moving branes has been given in \cite{V}, whose authors discussed their results with N. Seiberg. Reinforcing  conclusions from \cite{K}, the authors \cite{V} emphasize in addition that, within the approach with moving branes, no reasons are seen for the equivalence of the direct and dual ${\cal N}=1$ theories outside the conformal window.\\

The purpose of this paper is to introduce and to use the definite dynamical scenario which allows to calculate the mass spectra in ${\cal N}=1$ SQCD-like theories. The main dynamical assumption of this scenario  is that quarks in such theories can be in two different {\it standard} phases only\,: this is either the HQ (heavy quark) phase where they are confined, or the Higgs phase where they form nonzero coherent condensate breaking the color symmetry (and so not confined), at the appropriate values of the lagrangian parameters. The word "standard" implies also that, in addition to the ordinary mass spectrum described below in this paper, in such ${\cal N}=1$ theories without colored adjoint scalars (unlike the very special ${\cal N}=2$ SQCD with its additional colored scalar fields $X^{adj}$ and enhanced supersymmetry) no parametrically lighter solitons (e.g. magnetic monopoles or dyons, or others) are formed at those scales where the massless regime is broken explicitly by nonzero particle masses. (It is worth noting that the appearance of additional light solitons will influence the 't Hooft triangles of this  ${\cal N}=1$ theory). Let us emphasize also that this dynamical scenario satisfies all those tests which were used as checks of the Seiberg hypothesis about the equivalence of the direct and dual theories. This shows, in particular, that all these tests, although necessary, may well be insufficient.\\

The organization of this paper is the following. The direct and dual theories for quarks of equal small masses, $0<m_Q\ll \la$, are considered in sections 2-4 and 7. Other sections deal with quarks of unequal masses, when there are $\nl$ lighter flavors with masses $\ml$ and $\nh=(N_F-\nl)$ heavier flavors with masses $0<\ml<\mh\ll\la$. The mass spectra of both the direct and dual theories in the conformal window are described in Sections 2-5. It is shown that in all cases where an additional small parameter is available (it is $0<\bo/N_F=(3N_c-N_F)/N_F\ll 1$ at the right end of the conformal window and its dual analog $0<\bd/N_F=(2N_F-3N_c)/N_F\ll 1$ at the left end), the parametrical differences in the mass spectra of the direct and dual theories can be traced. Sections 6-9 deal with the direct and dual theories in some special regimes of interest. Finally, some conclusions are presented in Section 10. Besides, there is one Appendix.

\section{ Direct theory.  Equal quark masses. {\boldmath $0<\bo/N_F\ll 1$}}

The Lagrangian of the direct theory at scales $\mu>\la$ has the form\,:
\bbq
K={\rm Tr}\Bigl ( Q^\dagger Q + {\ov Q}^\dagger{\ov Q}\Bigr )\,,\quad
{\w}= -\frac{2\pi}{\alpha(\mu)} S+{\rm Tr}\,\bigl ( m_Q(\mu)\,{\ov Q} Q\bigr )\,,\quad S= -W_{\alpha}^2/32\pi^2\,.
\eeq
Here\,:\, $\alpha(\mu)$ is the running gauge coupling (with its scale parameter $\la$\, {\it independent
of quark masses}),\,${\rm  W}_{\alpha}$ is the gluon field strength, and $m_Q(\mu)\ll \la$ is the running quark mass.
\footnote{\,
The gluon exponents in Kahler terms are implied here and everywhere below.

Besides, everywhere below\,: $A\approx B$ has to be understood as an equality neglecting smaller power corrections, and $A\ll B$ has to be understood as $|A|\ll |B|$. \label{(f1)}
}

The theory is UV free and is in the conformal regime at scales $\mu_H<\mu<\la$
\footnote{\,
By definition, within the conformal window, $\la$ is a scale such that the coupling $a(\mu=\la)=N_c\alpha(\mu=\la)/2\pi$ is sufficiently close to its fixed point value $a_{*}$, i.e. $a_{*}-a(\mu=\la)=\delta\, a_{*}\,,\,\, \delta\ll 1$ (and similarly for the dual theory, ${\ov a}_{*}-{\ov a}(\mu=\Lambda_q)=\delta\, {\ov a}_{*}$\,). The coupling ${\it a}_*$ is weak at $\bo/N_F\ll 1$, $\gamma_Q(a_*)=(\bo/N_F)\simeq (1-1/N_c^{2}){\it a}_*\simeq {\it a}_* \ll 1$\,. Here and in what follows, we trace only the leading exponential dependence on the small parameter $\bo/N_F\ll 1$ (or $\bd/N_F\ll 1$)\,, i.e. factors of the order of $\exp\{-c_o(N_F/\bo)\},\, c_o=O(1)$, while the nonleading terms of the order of $\exp\{-c_o\delta(N_F/\bo)\}$ or pre-exponential factors $\sim (N_c/\bo) ^{\sigma},\, \sigma=O(1)$\,, are neglected, because this simplifies greatly all expressions. Besides, it is always implied that even when $\bo/N_F$ or $\bd/N_F$ are $\ll 1$, these small numbers do not compete {\it in any way} with the main small parameter $m_Q/\la$.  See the Appendix for more details.
}
, where $\mu_H$ is the highest physical mass, and in this case it is the quark pole mass $\mq$\,:
\bbq
\frac{\mq}{\la}=\frac{m_Q(\mu=\mq)}{\la}=\frac{m_Q}{z_Q(\la,\mq)\la}\sim\frac{m_Q}{\la}\Biggl (\frac
{\la}{\mq}\Biggr )^{\gamma_Q}\sim\Biggl (\frac{m_Q}{\la}\Biggr )^{\frac{1}{1+\gamma_Q}=
\frac{N_F}{3N_c}}\,,
\eeq
\bq
m_Q\equiv m_Q(\mu=\la)\,, \quad\gamma_Q=\frac{\bo}{N_F},\quad \bo=3N_c-N_F\,, \quad z_Q(\la,\mq)\sim \Bigl (\frac{\mq}{\la}\Bigr )^{\gamma_Q}\,, \label{(2.1)}
\eq
where $z_Q(\la,\mq)$ is the renormalization factor of the quark Kahler term and $\gamma_Q$ is the quark anomalous dimension \cite{NSVZ}. After integrating out
all quarks as heavy ones, a pure Yang-Mills (YM) theory with $N_c$ colors remains
at lower energies.  Its scale parameter $\lym$ can be determined from matching the couplings of
the higher- and lower-energy theories at $\mu=\mq$. Proceeding as in \cite{ch1}, we obtains
\bq
3 N_c\ln\Biggl (\frac{\mq}{\lym} \Biggr )\approx \frac{2\pi}{\alpha_*}\approx N_c\frac{N_F}{\bo}
\quad \ra \quad\lym\sim\exp \Bigl \{\frac{-N_F}{3\bo}\Bigr \}\,\mq \ll m_Q^{\rm pole}\,.\label{(2.2)}
\eq

Therefore, there are two parametrically different scales in the mass spectrum of the direct
theory in this case. There is a large number of colorless flavored hadrons made of
weakly confined (the string tension $\sqrt \sigma\sim \lym\ll m_Q^{\rm pole}$) and weakly
interacting {\it nonrelativistic} heavy quarks $Q\,,\,\ov Q$ with masses $m_Q^{\rm pole}\gg \lym$\,,
and a large number of gluonia with the mass scale $\lym=\exp \{-N_F/3\bo\}m_Q^{\rm pole}\ll
m_Q^{\rm pole}$.

To check the self-consistency, we estimate the scale of the gluon masses due to a possible
quark higgsing. The quark chiral condensate $\cm^2$ at $\mu=\la$ is given by \cite{Konishi}\,:
\bq
\frac{\cm^2}{\la^2}\equiv \frac{1}{\la^2}\langle 0|{\ov Q}Q (\mu=\la) |0\rangle =\frac{\langle S
\rangle=\lym^3}{m_Q\la^2}\sim\exp \Bigl \{\frac{-N_F}{\bo} \Bigr \}\Biggl (\frac{m_Q}{\la}\Biggr )
^{\frac{N_F-N_c}{N_c}}\,.\label{(2.3)}
\eq
If the gluons were acquired masses  $\mg > m_Q^{\rm pole}$ due to higgsing of quarks, then the conformal renormalization group (RG) evolution would stops at $\mu=\mg$. Hence, $\mg$ can be estimated from
\bbq
\mg^2\sim a(\mu=\mg)\langle 0|{\ov Q}Q (\mu=\mg)|0\rangle\,,\quad
\mg^2\sim \cm^2\Biggl (\frac{\mg}{\la} \Biggr )^{\gamma_Q}\,,\quad a(\mu)\equiv N_c g^2(\mu)/8\pi^2=
\frac{N_c\alpha(\mu)}{2\pi}\,,
\eeq
\bq
\frac{\mg}{\la}\sim \Biggl (\frac{\cm^2}{\la^2} \Biggr )^{\frac{1}{2-\gamma_Q}}\sim
\exp \Bigl \{\frac{-N_c}{2\bo} \Bigr \}\frac{\lym}{\la}\ll \frac{\lym}{\la}\ll \frac
{m_Q^{\rm pole}}{\la}\,.\label{(2.4)}
\eq
Therefore, the scale of possible gluon masses, $\mu=\mg$, is parametrically smaller not only than
the quark pole mass but also than $\lym$, $\mg\ll\lym\ll m_Q^{\rm pole}$, and the picture of $Q\,,\ov Q$ being in the HQ (heavy quark) phase is self-consistent.

\section{ Dual theory. Equal quark masses. {\boldmath $0<\bo/N_F\ll 1$}}

The Lagrangian of the dual theory at the scale $\mu=|\Lambda_q|$ is taken in the form \cite{S1}
\bq
{\ov K}= {\rm Tr}\Bigl ( q^\dagger q + {\ov q}^\dagger
{\ov q}\Bigr )+\frac{1}{\mu_2^2}{\rm Tr \Bigl (M^{\dagger}M\Bigr )}\,,\label{(3.1)}
\eq
\bbq
{\ov \w}=\Biggl [\,\, -\frac{2\pi}{\ov \alpha(\mu)}\, {\ov S}+\frac{1}{\mu_1}\,\rm {Tr} \Bigl ({\ov q}\,
M\, q \Bigr ) + {\rm Tr \Bigl (\,{\ov m}_Q(\mu) M}\Bigr )\,\Biggr ]\,,
\quad {\ov S}=-{\ov W}_{\alpha}^2/32\pi^2\,.
\eeq
Here\,:\, the number of dual colors is ${\ov N}_c=(N_F-N_c),\, \bd=3\nd-N_F$, and $M_{ij}$ are the $N_F^2$ elementary mion fields, ${\ov a}(\mu)=\nd{\ov \alpha}(\mu)/2\pi=\nd{\ov g}^2(\mu)/8\pi^2$ is the dual running gauge coupling,\,\, ${\rm \ov W}^b_{\alpha}$ is the dual gluon field strength. The gluino condensates of the direct and dual theories are matched, $\langle{-\,\ov S}\rangle=\langle S\rangle=\lym^3$.

By definition, $\mu=|\Lambda_q|$ is such a scale that the dual theory already entered sufficiently deep into the conformal regime, i.e. both the gauge and Yukawa couplings, ${\ov a}(\mu=|\Lambda_q|)$ and
$a_f(\mu=|\Lambda_q|)=\nd f^2(\mu=|\Lambda_q|)/8\pi^2,\,\,f(\mu=|\Lambda_q|)=\mu_2/\mu_1$, are already close to their fixed point values, $\ov\delta=({\ov a}_*-{\ov a}(\mu=|\Lambda_q|)/{\ov a}_*\ll 1$ (and similarly for $a_f(\mu=|\Lambda|_q)$\,. At $0<\bo/N_c\ll 1$ the fixed point dual couplings are ${\ov a}_*\sim a^*_f\sim 1$, while at $0<\bd/\nd\ll 1$ they are small, ${\ov a}_*\sim a^*_f\sim \bd/\nd\ll 1$\,.

We take $|\Lambda_q|=\la$ for simplicity (because this does not matter finally but simplifies greatly all formulas, see the Appendix for more details).
~\footnote {\,
The phase of ${\ov\Lambda}_{YM}$ has to be chosen appropriately \cite{IS} to ensure $\langle{-\,\ov S}\rangle=\langle S\rangle$. This is always implied in what follows.
}
The condensates $\langle M_{ji}(\mu=\la)\rangle$ and $\langle{\ov Q}_j Q_i(\mu=\la)\rangle$ can always
be matched at $\mu=|\Lambda_q|=\la$, at the appropriate choice of $\mu_1$ in \eqref{(3.1)}, see below,
$$M_o\equiv\langle 0|M(\mu=\la)|0\rangle=\langle 0|{\ov Q}Q (\mu=\la) |0\rangle=\cm^2\,.$$

Because the gluino condensates are also matched, it follows from the Konishi anomalies \cite{Konishi} that
$$\langle S\rangle=\lym^3=m_Q\cm^2=|\langle {\ov S}\rangle|= |{\ov \Lambda}_{YM}|^3={\ov m}_Q(\mu=\la) M_o\,,\quad {\ov m}_Q(\mu=\la)=m_Q\,.$$

At $3/2<N_F/N_c<3$ this dual theory can be taken as UV free at $\mu\gg\la$ and this requires that its Yukawa coupling at $\mu=\la,\, f(\mu=\la)=\mu_2/\mu_1$, cannot be larger than its gauge coupling ${\ov g}(\mu=\la)$, i.e. $\mu_2/\mu_1\lesssim 1$. The same requirement to $\mu_2/\mu_1$ follows from the conformal behavior of this theory at $3/2<N_F/N_c<3$ and $\mu<\la$, i.e. $a_f(\mu=\la)\simeq a^*_f\sim {\ov a}_*=O(\bd/\nd)$. Hence, we take $\mu_2=\mu_1$ in what follows. Therefore, only one free parameter $\mu_1\equiv Z_q \la$ remains in the dual theory. It will be determined below from the explicit matching of the gluino condensates in the direct and dual theories. We consider below this dual theory at $\mu\leq \la$ only where it claims to be equivalent to the direct one.\\

The dual theory is also in the conformal regime at $0<\bo/N_F\ll 1$ and ${\ov \mu}_H<\mu<\la$\,, where
${\ov \mu}_H$ is the corresponding largest physical mass.

Assuming that the dual quarks ${\ov q},\,q$ are in the HQ-phase and hence ${\ov \mu}_H=\mu_q^{\rm pole}$\,,
we find their pole mass ($\nd=N_F-N_c\,,\,\,\bd=3\nd-N_F\,,\,\, \gamma_q=\bd/N_F$\,)\,:
\bq
\frac{\mu_q^{\rm pole}}{\la}\sim\frac{\mu_q}{\la} \Biggl (\frac{\la}{\mu_q^{\rm pole}}\Biggr )^
{\gamma_q}\sim\Biggl (\frac{\cm^2}{Z_q\la^2}\Biggr )^{N_F/3\nd}\,,\quad \mu_q\equiv\mu_q(\mu=\la)=
\frac{M_o}{\mu_1}=\frac{\cm^2}{Z_q\la}\,.\label{(3.2)}
\eq

We now integrate out all dual quarks as heavy ones at $\mu<\mu_q^{\rm pole}$ and determine the scale factor ${\ov\Lambda}_{YM}$ of the remained dual YM theory (the dual coupling is ${\ov  a_*}=\nd\,{\ov
\alpha_*}/2\pi=O(1)$\,)\,:
\bq
3\nd\ln\Biggl (\frac{\mu_q^{\rm pole}}{{-\ov \Lambda}_{YM}} \Biggr )\sim \frac{\nd}{\ov a_*}
\sim \nd \quad \ra \quad {-\ov \Lambda}_{YM} \sim\mu_q^{\rm pole}\,.\label{(3.3)}
\eq
Equating the gluino condensates of the direct and dual theories, $(-{\ov \Lambda}_{YM})^3=\lym^3$,
we obtain
\bq
{-\ov \Lambda}_{YM}=\lym \quad \ra \quad Z_q\sim\exp \Bigl \{\frac{-N_F}{3\bo}\Bigr \}\ll 1\,.\label{(3.4)}
\eq

We now find the mass $\mu_M$ of the dual mesons $M$ (mions). It can be found from their
effective Lagrangian, obtained by integrating out all dual quarks and gluons. Proceeding as
in \cite{ch1} and integrating out all dual quarks as heavy ones and all dual gluons via
the Veneziano-Yankielowicz (VY) procedure \cite{VY}, we obtain the Lagrangian of mions\,
(the fields $M$ are normalized in \eqref{(3.5)} and \eqref{(3.6)} at $\mu=\la$ and $\bd=3\nd-N_F$)
\bq
K_M=\frac{z_M(\la,\,\mu_q^{\rm pole})}{Z_q^2\la^2}\,{\rm Tr\,}(M^{\dagger}M)\,,\quad
\w_M= -\nd \Biggl ( \frac{\det M}{Z_q^{N_F}\la^{\bo}}\Biggr )^{1/\nd}+ {\rm Tr\,} (m_Q M)\,,\label{(3.5)}
\eq
\bbq
z_M=z_M(\la,\,\mu_q^{\rm pole})\sim\Biggl (\frac{\la}{\mu_q^{\rm pole}}\Biggr )^{2\gamma_q}\sim\Biggl
(\frac{\la}{\lym} \Biggr )^{2\,\bd/N_F}\gg 1\,,\quad m_Q\langle M\rangle=m_Q\cm^2=\langle S\rangle=\lym^3\,.
\eeq
In a number of cases, it is also convenient to write the superpotential of mions in the form
\bq
\w_M= -\nd\, \lym^3\Biggl (\det \frac{ M}{\langle M \rangle}\Biggr )^{1/\nd}+ {\rm Tr\,}
(m_Q M)\,.\label{(3.6)}
\eq

It follows from \eqref{(3.5)} that
\bq
\mu_M\sim\frac{Z_q^2\la^2}{z_M}\,\frac{\langle S\rangle}{\langle M\rangle^2}\sim \frac{Z_q^2 m_Q\la^2}{z_M \langle M\rangle }\sim \frac{Z_q^2 m_Q\la^2}{z_M\cm^2}\sim\lym\,.\label{(3.7)}
\eq

To check that the above assumption that the dual quarks $q,\,\ov q$ are in the HQ phase (i.e. are not higgsed) is not self-contradictory, we estimate the mass ${\ov \mu}_{gl}$ of dual gluons due to possible
higgsing of the dual quarks. The condensate of dual quarks at $\mu=\la$ is \cite{Konishi}\,:
\bq
\mu_o^2\equiv \mu_C^2(\mu=\la)=|\langle {\ov q}q (\mu=\la)\rangle |=\mu_1 m_Q=Z_q\la m_Q\,,
\quad \mu_C^2(\mu<\la)\sim\mu_o^2\Biggl (\frac{\mu}{\la} \Biggr )^{\gamma_q}\,.\label{(3.8)}
\eq
Therefore, the mass of dual gluons due to possible higgsing of dual quarks can be estimated as
\bq
{\ov \mu}^2_{gl}\sim \Biggl (\ov a_*\sim 1 \Biggr )\Biggl (Z_q\la m_Q \Biggr )\Biggl (\frac
{{\ov \mu}_{gl}}{\la}\Biggr )^{\bd/N_F}\, \quad \ra \quad {\ov \mu}_{gl} \sim \lym\,.\label{(3.9)}
\eq
It is seen that this is, at least, not parametrically larger than the pole mass of the dual
quark,\, ${\ov \mu}_{gl} \sim \mu_q^{\rm pole}$. Hence, it may actually be even somewhat smaller, i.e.
${\ov \mu}_{gl}=\mu_q^{\rm pole}/(\rm several)$ (because we do not trace
non-parametrical factors $O(1)$), and the dual quarks are indeed in the HQ-phase. In
similar situations here and everywhere below we will assume that this is indeed the case.
\footnote{\,
Due to the rank restriction at $N_F>N_c$, the opposite case in which quarks were actually higgsed would result in a spontaneous breaking of the global flavor symmetry (and breaking of the whole gauge group) and in appearance of a large number of strictly massless Nambu-Goldstone particles. In what follows, in favor of Seiberg's duality, it is assumed that, as in the direct theory, the global flavor symmetry is not broken spontaneously in the dual theory also, and so this variant is not realized. The same arguments about the absence of spontaneous breaking of corresponding global flavor symmetries and absence of strictly massless Nambu-Goldstone particles are applicable to both direct and dual theories everywhere below in the text. \label{(f4)}
}

On the whole, there is only one scale $\sim \lym$ in the mass spectrum of the dual theory in
this case. The masses of dual quarks $q,\,\ov q$, mions $M$ and dual gluonia are all
$\sim \lym$.\\

Comparing the mass spectra of the direct and dual theories it is seen that they are
parametrically different. The direct quarks have large pole masses $m_Q^{\rm pole}/\lym\sim
\exp\{N_F/3\bo\}\gg 1$ and are parametrically weakly coupled and nonrelativistic inside
hadrons (and weakly confined, the string tension ${\sqrt \sigma}\sim \lym\ll m_Q^
{\rm pole}$\,), and therefore the mass spectrum of low-lying mesons is Coulomb-like, with small
mass differences $\delta \mu_H/\mu_H=O(\bo^2/N_F^2)\ll 1$ between nearest hadrons.
All low-lying gluonia have masses $\sim \lym$, these are parametrically smaller than the masses of
flavored hadrons. At the same time, all hadron masses in the dual theory are of the same scale
$\sim \lym$, and all couplings are $O(1)$, and hence there is no reason for parametrically small mass
differences between hadrons. We conclude that the direct and dual theories are not equivalent.

\section{ Direct and dual theories. Equal quark masses,\\ {\boldmath $0<\bd/N_F\ll 1$}}

This case is considered analogously to those in Sections 2 and 3, and we therefore highlight
only on differences from Sections 2 and 3. The main difference is that at $\mu\sim \la$, the direct
coupling is not small now, $a_*=O(1)$, while both the gauge and Yukawa couplings of the dual theory are
parametrically small,\, ${\ov a}_*\sim a_f^{*}\sim\bd/N_F\ll 1$.

The pole mass of direct quarks, $m_Q^{\rm pole}$, is the same as in \eqref{(2.1)}. But $\lym$ is now parametrically the same\,: $\lym\sim m_Q^{\rm pole}$ (as is the scale of the gluon mass due to possible quark higgsing, $\mg\sim m_Q^{\rm pole}\sim \lym$). Assuming that quarks are not higgsed but are in the HQ-phase (see the footnote \ref{(f4)}), one obtains that there is only one mass scale $\mu_H\sim \lym \sim \la(m_Q/\la)^{N_F/3N_c}$ in the mass spectrum of hadrons in the direct theory in this case.\\

We now consider the weakly coupled dual theory and recall that, by definition, $\mu=|\Lambda_q|$ is such a scale that, at $0<\bd/\nd\ll 1$, the dual theory at $\mu=|\Lambda_q|$ already entered sufficiently deep into the conformal regime. I.e., both the gauge and Yukawa couplings, ${\ov a}(\mu=|\Lambda_q|)$ and $a_f(\mu=|\Lambda_q|)$, are already close to their small fixed point values, $\ov\delta=({\ov a}_*-{\ov a}(\mu=|\Lambda_q|)/{\ov a}_*\ll 1$, and similarly for $a_f(\mu=|\Lambda_q|)$.  We take $|\Lambda_q|=\la$
everywhere below for simplicity (because this does not matter finally but simplifies greatly all formulas, see the Appendix for more details).

Hence, the pole mass of dual quarks looks now as
\bq
\frac{\mu_q^{\rm pole}}{\la}\sim\Biggl (\frac{\mu_q}{\la} \Biggr )^{N_F/3\nd}\,, \quad
\frac{\mu_q= \mu_q(\mu=\la)}{\la}=\frac{\cm^2}{Z_q \la^2}\,,\quad \frac{\cm^2}{\la^2}
=\frac{\lym^3}{m_Q\la^2}=\Biggl (\frac{m_Q}{\la}\Biggr )^{\nd/N_c}\,. \label{(4.1)}
\eq

After integrating out all dual quarks as heavy ones at $\mu=\mu_q^{\rm pole}$\,,
the dual YM theory remains (together with the mions $M$). The scale parameter ${\ov \Lambda}=\langle {\ov
\Lambda}_L(M)\rangle$ of its gauge coupling can be found from \cite{KSV} \,:
\bq
3\nd \ln\Biggl ( \frac{\mu_q^{\rm pole}}{{-\ov \Lambda}}\Biggr )\approx \frac{2\pi}{\ov
\alpha_*}\approx \frac{N_F}{\bd}\,\frac{\nd^2-1}{2N_F+\nd}\approx \frac{3\nd^2}{7\,\bd}\,.\label{(4.2)}
\eq

Now, from matching the gluino condensates in the direct and dual theories, we obtain
\bq
|\ov \Lambda|=\lym \quad \ra \quad  Z_q\sim\exp \Bigl \{\frac{-\nd}{7\,\bd} \Bigr \}
\ll 1\,,\quad \mu_q^{\rm pole}\sim\exp \Bigl \{\frac{\nd}{7\,\bd} \Bigr \}\lym \gg \lym\,.\label{(4.3)}
\eq
\vspace{3mm}

On the whole, the expressions for $\lym$ and $Z_q$ can be written in the general case where $\bo>0,\,\bd>0$ as
\bq
\frac{\lym}{\la} \sim \exp\Biggl \{-\frac{N_c}{\bo} \Biggr \}\Biggl ( \frac{\det m_Q}{\la^{N_F}} \Biggr )^{1/3N_c}\,,\quad
Z_q\sim  \exp \Biggl \{-\Biggl (\frac{N_c}{\bo}+\frac{\nd}{7\,\bd}\Biggr ) \Biggr \}\,. \label{(4.4)}
\eq
The symbol $\sim$ in \eqref{(4.4)} denotes the exponential accuracy in dependence on the large parameters $N_c/\bo\gg 1$ or $\nd/\bd\gg 1$. Hence,
if $N_c/\bo$ or $\nd/\bd$ are $O(1)$, then the dependence on these has to be omitted from \eqref{(4.4)}. For our purposes, this
exponential accuracy in \eqref{(4.4)} and below will be sufficient. \\

Therefore, similarly to the case of the weakly coupled direct theory in section 2, the dual
quark pole mass $\mu_q^{\rm pole}$ is parametrically larger here than $\lym$.

Then, proceeding as in \cite{ch1} and integrating out the dual gluons via the VY-procedure yields
the Lagrangian of mions $M$\,, which can be written in the form \eqref{(3.6)}. The mion masses are therefore given by
\bq
\mu_M\sim m_Q\Biggl (\frac{\mu_2^2}{z_M\cm^2} \Biggr )\sim m_Q\Biggl (\frac{{ Z_q}^2\la^2}{z_M\cm^2} \Biggr ),\,\,
z_M=z_M(\la,\,\mu_q^{\rm pole})\sim \Biggl (\frac{\la}{\mu_q^{\rm pole}}\Biggr )^{2\bd/N_F}\gg 1\,. \label{(4.5)}
\eq
It follows from \eqref{(4.5)} that
\bq
\mu_M\sim { Z_q}^2 \lym=\exp \Bigl \{ \frac{-2\nd}{7\,\bd}\Bigr \}\lym \ll \lym\,. \label{(4.6)}
\eq
Therefore, the mion masses are parametrically smaller than $\lym$.

To check that there are no self-contradictions and the dual quarks are not higgsed, it remains to estimate the gluon masses due
to  possible higgsing of these dual quarks. We have
\bq
\frac{{\ov \mu}^{2}_{gl}}{\la^2}\sim { Z_q}\frac{m_Q}{\la}\Biggl ( \frac{{\ov \mu}_{gl}}{\la}\Biggr )^{\bd/N_F}\, \quad \ra \quad {\ov \mu}_{gl}\sim \exp \Bigl \{\frac{-\nd}{14\,\bd} \Bigr \}\lym\ll \lym \ll\mu_q^{\rm pole}\,. \label{(4.7)}
\eq
\vspace{3mm}

Therefore, there are three parametrically different mass scales in the dual theory in this  case.\\
\,\, a)\, A large number of flavored hadrons made of weakly coupled nonrelativistic
(and weakly confined, the string tension $\sqrt \sigma\sim \lym \ll \mu_q^{\rm pole}$) dual
quarks with the pole masses $\mu_q^{\rm pole}/\lym=\exp\{\,\nd/7\,\bd\}\gg 1$. The mass spectrum of
low lying flavored mesons is Coulomb-like, with parametrically small mass differences $\delta \mu_H/
\mu_H=O(\bd^2/N^2_F)\ll 1$.\\
\,\, b)\, A large number of gluonia with the mass scale $\sim \lym$.\\
\,\,c)\, $N_F^2$ lightest mions $M$ with parametrically smaller masses, $\mu_M/\lym\sim \exp
\{-2\nd/7\,\bd \}\ll 1$.

At the same time, there is only one mass scale $\sim \lym$ of all
hadron masses in the direct theory which is strongly coupled here, $a_*=O(1)$. Clearly, the mass
spectra of the direct and dual theories are parametrically different.\\

On the whole, it follows that when an appropriate small parameter is available ($0<\bo/
N_F\ll 1$ when the direct theory is weakly coupled, or $0<\bd/N_F\ll 1$ when weakly coupled is
the dual theory), the mass spectra of the direct and dual theories are parametrically
different. Therefore, there are no reasons for these mass spectra to become exactly the same when
$\bo/N_F$ and $\bd/N_F$ become $O(1)$.

\section{ Direct and dual theories. Unequal quark masses, \\ {\boldmath $ 0<\bd/N_F\ll 1\,,\,\,N_c<\nl<3N_c/2$}}

The standard consideration that allows "to  verify" that the duality works properly for quarks of unequal masses is as follows \cite{S1, IS}. For instance, we take $\nh$ quarks of the direct theory to be heavier than the other $N_c<\nl=N_F-\nh$ quarks. Then, after integrating out these
$\rm h$-flavored quarks as heavy ones, the direct theory with $N_c$ colors and $\nl$ of $\it l$-flavors remains at lower energies. By duality, it is equivalent to the dual theory with $\nl-N_c$ colors and $\nl$ flavors. On the other side, the original theory is equivalent to the dual theory with $\nd=N_F-N_c$ colors and $N_F$ flavors. In this dual theory, the $\rm h$-flavored dual quarks are assumed to be higgsed, and hence the dual theory with $(N_F-N_c-\nh)=\nl-N_c$ colors and $N_F-\nh=\nl$ of $\it l$-flavors remains at lower energies. Therefore, all looks self-consistent.\\

But we consider now this variant in more details, starting from the left end of the conformal window, $0<\bd/N_F\ll 1$\,. At $\mu < \la$, the dual theory, by definition, has already entered sufficiently deep the conformal regime and therefore both its gauge and Yukawa couplings are close to their parametrically small frozen values, ${\ov a}_*\sim a_f^*\sim \bd/N_F\ll 1$. We take $N_c<\nl<3N_c/2$ for the direct quarks $\ql\,,\oql$ to have smaller masses $\ml$ at $\mu=\la$\,, and the other $\nh=N_F-\nl$ quarks $\qh\,,\, \oqh$ to have larger masses $\mh\,, \,\,r\equiv \ml/\mh < 1$.\\

\hspace*{1cm}{\bf 5.1. Direct theory} \\

We start from the direct theory, because, in a sense, its mass spectrum is easier to calculate. It is in the conformal regime at $\mu<\la$\,, and the highest physical mass scale $\mu_H$ is here the pole mass of the heavier $\qh\,,\,\oqh$ quarks\,:
\bq
\frac{\mhp}{\la}\sim\frac{\mh}{\la}\,\Biggl (\frac{\la}{\mhp}\Biggr )^{\gamma_Q=\bo/N_F}\sim\Biggl (\frac
{\mh}{\la}\Biggr )^{N_F/3N_c}\,. \label{(5.1)}
\eq

After integrating these quarks out, the lower energy direct theory with $N_c$ colors and
$N_c<\nl<3N_c/2$ flavors of lighter $\ql\,,\,\oql$ quarks remains.
\footnote{\,
To simplify all formulas, the value $(3N_c-2\nl)/\nl$ is considered as $O(1)$ quantity in this section,
 }
From matching with the coupling $a_{*}=O(1)$ of the higher energy theory, its gauge coupling is also $O(1)$ at $\mu=\mhp$ and therefore the scale parameter  of this gauge coupling is $\la^{\prime}\sim\mhp$. The current masses of $\ql\,,\,\oql$ quarks at $\mu=\mhp$ are $\hat \ml\equiv
\ml(\mu=\mhp)=r\,\mhp\ll \la^{\prime}=\mhp,\,\, r\ll 1$.

Now, how to deal further with this theory at lower energies ?\, As was described in \cite{ch1} (see section 7 therein), there are two variants, "a" and "b".  All hadron masses $\mu_{H}$ of the lower energy theory are much smaller than $\la^{\prime}$ in the variant "a", $\mu_H\ll \la^{\prime}$. At $\mu<\la^{\prime}$ this theory enters the strong coupling regime with $a(\mu\ll\la^{\prime})\gg 1$. Its mass spectrum is, in essence, the same as those described below in section 6 (and only the value of $N_F$ is different). And this mass spectrum is parametrically different from those of the original dual theory (see below in this section).

We therefore do not consider this variant "a" in this section, because our purpose here is to check the duality in the variant most favorable for its validity. This is the variant "b"\,: "confinement without chiral symmetry breaking" (although some general arguments  have been presented in section 7 of \cite{ch1} that this variant cannot be realized). This amounts {\it to assuming} (because all original direct degrees of freedom cannot "dissolve in the pure air") that due to strong non-perturbative confining effects, the direct quarks $\ql\,,\oql$ and gluons form a large
number of heavy hadrons with masses $\mu_H\sim\la^{\prime}$.

Instead, new light composite particles (special solitons) appear whose masses are parametrically smaller than $\la^{\prime}$. These are the dual
quarks ${\hat q}^{\it l},\, {\hat{\ov q}}_{\it \ov l}$, the dual gluons with $\nd^{\,\prime}=\nl- N_c$ of dual colors, and the mions ${\hat M}_{\it l}
\equiv {\hat M}_{l}^{\ov l}$. Their interactions at scales $\mu<\la^{\prime}$ are described by the Seiberg dual Lagrangian. Their dual gauge coupling ${\ov a}(\mu)$ at $\mu\lesssim \la^\prime\sim \mhp$ is ${\ov a}(\mu=\la^\prime/(\rm several))\sim {\it a}^*=O(1)$,
and therefore the scale parameter of this dual gauge coupling is $\sim \la^{\prime}\sim \mhp$.

The dual Lagrangian (at the scale $\mu\sim \la^{\prime}\,,\, (\cma)^2\equiv \langle {\ov Q}^{\it l}
Q_{\it l}(\mu=\la)\rangle=\langle S\rangle/\ml =\lym^3/\ml $\,) is
\bq
\ov K={\rm Tr\,}_{\it l} \bigl( {\hat q}^{\dagger} {\hat q}
+ {\hat {\ov q}}^{\dagger} {\hat {\ov q}}\bigr )+\frac{1}{(\la^{\prime})^2}
{\rm Tr} \bigl ({\hat M}_{l}^{\dagger}{\hat M}_{l}\bigr )\,, \label{(5.2)}
\eq
\bbq
{\ov\w}= -\frac{2\pi}{\ov \alpha(\la^{\,\prime})}\, {\ov S}+\frac{1}{\la^{\,\prime}}\,{\rm Tr}_{\it l}
\Bigl ({\hat {\ov q}}\,{\hat M}_{\it l}\, {\hat q} \Bigr ) +{\hat \ml}\,{\rm Tr} \, {\hat M}_
{\it l}\,;\quad  {\ov S}=-{\ov W}_{\alpha}^2/32\pi^2\,,\quad \langle
{\hat{\ov q}}_{\it \ov l}\, {\hat q}^{\it l}\rangle=-\delta^{\it l}_{\it \ov l}\,{\hat m}_{\it l}
\,{\la^{\,\prime}}\,,
\eeq
\bq
\hml\equiv\ml(\mu=\la^{\,\prime})=\ml\,z_Q^{-1},\,\, \langle {\hat M}_{\it l} \rangle=(\cma)^2\,z_Q,\,\, z_Q\sim
\Biggl (\frac{\mhp}{\la}\Biggr )^{\gamma_Q=\frac{\bo}{N_F}}\sim\Biggl (\frac{\mh}{\la}\Biggr )^{\frac{\bo}{3N_c}}\ll 1\,. \label{(5.3)}
\eq

This dual theory is in the IR free weakly coupled HQ-phase at $\mu<\la^{\prime}$. At lower scales $\mulp<\mu<\la^
{\,\prime}$ both its gauge and Yukawa couplings decrease logarithmically (\, $\bd^{\,\prime}=3\nd^{\,\prime}-
\nl <0$) and become much less than unity at $\mu=\mulp\ll \la^{\,\prime}$ and $r\ll 1$. The pole mass $\mulp$
of the quarks ${\hat q}^{\it l},\, {\hat {\ov q}}_{\it \ov l}$ is
\bq
\mulp=\mu_{q,\,\it l}/z^{\,\prime}_q=\langle{\hat M}_{\it l}\rangle/z^{\,\prime}_q\la^{\,\prime}=
\Biggl (r=\frac{\ml}{\mh}\Biggr )^{\frac{\nl-N_c}{N_c}}\,\mhp/z^{\,\prime}_q\ll \mhp\,, \label{(5.4)}
\eq
where $z^{\,\prime}_q=z^{\,\prime}_q(\la^{\,\prime},\, \mulp)\ll 1$ is the perturbative
logarithmic renormalization factor of dual {\it l}\,-quarks\,:
\bbq
z^{\,\prime}_q=z^{\,\prime}_q(\la^{\,\prime},\, \mulp)\sim\Biggl(\frac{{\ov \alpha}(\mulp)}{{\ov \alpha}
(\la^{\,\prime})}\Biggr )^{\nd^{\,\prime}/|\bd^{\,\prime}|}\sim\Biggl(\frac{1}{\ln (\la^{\,\prime}/\mulp)} \Biggr )^{\frac{\nl-N_c}{3N_c-2\nl}}\ll 1\,.
\eeq

At scales $\mu<\mulp$ all quarks ${\hat q}^{\it l},\, {\hat {\ov q}}_{\it \ov l}$ can be integrated
out as heavy ones, and there remains the dual YM theory with $\nd^{\,\prime}=\nl-N_c$ colors and mions
${\hat M}_{\it l}$\,. The scale factor $\oym=\langle {\ov \Lambda}_L({\hat M}_{\it l})\rangle$
(with mions ${\hat M}_{\it l}$ sitting down on ${\ov \Lambda}_L({\hat M}_{\it l}$)\,) of its gauge
coupling is determined from the matching \cite{ch1,ch2}
\bbq
3\nd^{\,\prime} \ln \Biggl (\frac{\mulp}{-\oym} \Biggr )= \bd^{\,\prime} \ln \Biggl (\frac
{\mulp}{\la^{\,\prime}} \Biggr )+\nl\ln \Biggl (\frac{\mulp}{\mu_{q,\,\it l}} \Biggr )\quad\ra
\eeq
\bq
 -\oym =-\langle {\ov \Lambda}_L({\hat M}_{\it l})\rangle = \lym=\Biggl ( \la^{\bo}\ml^{\nl}\mh^{\nh}\Biggr )^{1/3N_c}\,.\label{(5.5)}
\eq
Applying now the VY procedure to integrate out dual gluons, we obtain the lowest-energy Lagrangian of mions ${\hat M}_{\it l}$\,:
\bq
K_M=\frac{z^{\,\prime}_M}{(\la^{\,\prime})^2}{\rm Tr} \Bigl ({\hat M}_{l}^{\dagger}{\hat M}_{l}\Bigr )\,\quad \w=\Biggl [\,\,-(\nl-N_c)\,
\Biggl (\,\frac{\det {\hat M}_{\it l}}{(\la^{\,\prime})^{(3N_c-\nl)}} \,\Biggr )^{1/(\nl-N_c)}
+\hat \ml\,{\rm Tr}\hat M_{\it l}\,\,\Biggr ]\,, \label{(5.6)}
\eq
where $z^{\,\prime}_M=z^{\,\prime}_M(\la^{\,\prime},\,\mulp)\gg 1$ is the perturbative logarithmic
renormalization factor of mions. From \eqref{(5.6)}, the masses of mions ${\hat M}_{\it l}$ are
\bq
\mu({\hat M}_{\it l})\sim \hat \ml\,\frac{(\la^{\,\prime})^2}{z^{\,\prime}_M\langle {\hat M}_{\it l}\rangle}
\sim \, \Biggl (r=\frac{\ml}{\mh}\Biggr )^{\frac{2N_c-\nl}{N_c}}\,\,\mhp/z^{\,\prime}_M\,,\quad r\ll 1\,,\label{(5.7)}
\eq
\bbq
\frac{\lym}{\mulp}\sim z^{\,\prime}_q\,\,r^{\Delta}\ll 1\,,\quad
\frac{\mu({\hat M}_{\it l})}{\lym}\sim \,r^{2\Delta}/z^{\,\prime}_M\,\ll 1\,,\quad
0<\Delta=\frac{3N_c-2\nl}{3N_c}<\frac{1}{3}\,.
\eeq

Therefore, on the whole for Section 5.1, when we started from the direct theory, then integrated out the heaviest $\qh\,, \oqh$ quarks at $\mu=\mhp\sim\lym(1/r)^{\nl/3N_c}\gg \lym$ and, in the variant "b" (i.e. "confinement without chiral symmetry breaking", see section 7 in \cite{ch1})
dualized the remained theory of the direct {\it l} - flavors and direct gluons, the mass spectrum is as follows.\\

{\bf 1)} There is a large number of heavy direct ${\it h}$\,-mesons $M^{\rm dir}_{\it h}$,
made of $\qh\,, {\ov Q}^{\it h}$ quarks (and/or antiquarks and direct gluons) with largest masses
$\sim \mhp\sim (1/r)^{\nl/3N_c}\lym\ll \la$.\\

{\bf 1')}\, In the variant "b" considered here, the dualization of $\ql\,,{\ov Q}^{\it \ov
l}$ quarks and direct gluons leaves behind a large number of heavy direct {\it l}\,-mesons
$M^{\rm dir}_{\it l}$ with different spins, made of $\ql\,, {\ov Q}^{\it \ov l}$ quarks
(and direct gluons), also with masses $\sim \mhp$.\,\, Besides, there is also a large
number of heavy  direct hybrid mesons $M^{\rm dir}_{\it lh}$\,, baryons $B_{\it l},\,B_{\it lh}$
and gluonia, all also with masses $\sim \mhp$\,.
All these heavy particles are strongly interacting\,, with  couplings $O(1)$\,.\\

All other particles are parametrically lighter and originate from the dual quarks ${\hat q}
^{\it l}\,,{\hat {\ov q}}_{\it \ov l}$\,, the dual gluons and mions ${\hat M}_{\it l}$.\\

{\bf 2)} There is a large number of {\it l}\,-flavored dual mesons and $b_{\it l}$ baryons made of
nonrelativistic (and weakly confined, the string tension $\sqrt \sigma\sim \lym
\ll \mulp$)\, ${\hat q}^{\it l}\,,{\hat {\ov q}}_{\it \ov l}$ \,\,- quarks, with their pole masses
$$\mulp\sim (r)^{(\nl-N_c)/N_c}\, \mhp/z^{\prime}_q\sim (1/r)^{(3N_c-2\nl)/3N_c}\lym/z^{\prime}_q\gg \lym\,.$$

{\bf 3)} There is a large number of gluonia with masses $\sim \lym\ll \mulp$.\\

{\bf 4)} The lightest are $\nl^2$ scalar mions ${\hat M}_{\it l}$ with masses
$\mu({\hat M}_{\it l})\sim (r )^{2\Delta}\,\lym/z^{\prime}_M\ll \lym\,.$ \\
\vspace{3mm}

\hspace*{1cm} {\bf 5.2\,\,  Dual theory} \\

 We return to the beginning of this section and start directly from the dual theory with $\nd$ dual colors,
$N_F$ dual quarks $q$ and $\ov q$, and $N_F^2$ mions $M_{i}^{\ov j}$. At the scale $\mu< \la$ the theory is already entered the weak coupling conformal regime, i.e. its coupling ${\ov a}(\mu)$ is close to ${\ov a}_*=\nd\,{\ov \alpha}_{*}/2\pi\approx 7\bd/3\nd \ll 1$, see footnote \ref{(f1)} and \eqref{(4.2)}. We first consider the case most favorable for the dual theory, where the parameter $r=\ml/\mh$ is already taken to be sufficiently small (see below). Then, in the scenario considered in this paper, the highest physical mass scale $\mu_H$ in the dual theory is determined by masses of dual gluons due to
higgsing of ${\ov q}_{\it h},\, q^{\it h}$ quarks\,:
$$\mu_H^2=\ogh^2\sim {\ov a}_{*}\langle{\ov q}_{\it h} q^{\it h}(\mu=\ogh)\rangle.$$
The mass spectrum of the dual theory in this phase can be obtained in a relatively standard way, similarly to \cite{ch1, ch2}. For this reason,
we will skip from now on some intermediate relations in similar situations in what follows.
The emphasis will be made on new elements that have not appeared before.\\

1)\,\,\, The masses of $(2\nd\nh-\nh^2)$ massive dual gluons and their scalar superpartners are
\bq
\Bigl ({\ogh}\Bigr )^2\sim \langle {\ov q}_{\it h} q^{\it h} \rangle \Biggl (\frac{\ogh}
{\la}\Biggr )^{\gamma_q}\sim \mu_1\mh\Biggl (\frac{\ogh}{\la} \Biggr )^{\gamma_q}
\sim Z_q\la\mh \Biggl (\frac{\ogh}{\la} \Biggr )^{\bd/N_F}\,,\label{(5.8)}
\eq
\bq
\frac{\ogh}{\la}\sim \exp\{-\nd/14\bd \}\, \Biggl (\,\frac{\mh}{\la} \Biggr )^{N_F/3N_c}\ll \Biggl (\,\frac{\mh}{\la} \Biggr )^{N_F/3N_c} \,,\quad {\it Z_q}
\sim \exp\{-\nd/7\bd \}\ll 1\,.\label{(5.9)}
\eq

2)\,\,\, The $\nl\nh$ hybrid mions $M_{\it hl}$ and $\nl\nh$ nions $N_{
\it lh}$ (these are those dual ${\it l}$\,-quarks that have higgsed colors, their partners $M_
{\it lh}$ and $ N_{\it hl}$ are implied and are not shown explicitly) can be treated
independently of other degrees of freedom, and their masses are determined mainly by their
common mass term in the superpotential,
\bq
K_{\rm hybr}\simeq{\hat z}_M\,{\rm Tr}\,\Biggl (\frac{M^{\dagger}_{\it hl}M_{\it hl} }{Z^2_q\la^2}\Biggr )+{\hat z}_q\,{\rm Tr}\,\Biggl (N^{\dagger}_{\it lh} N_{\it lh} \Biggr )\,,\quad {\ov\w}=\Bigl ( Z_q\mh\la\Bigr )^{1/2}{\rm Tr}\,\Biggl (\frac {M_{\it hl}N_{\it lh}}{Z_q\la}\Biggr )\,, \label{(5.10)}
\eq
where ${\hat z}_q$ and ${\hat z}_M$ are the perturbative renormalization factors of dual
quarks and mions,
\bbq
{\hat z}_q={\hat z}_q(\la,\,\ogh)\sim\Biggl (\frac{\ogh}{\la} \Biggr )^{\bd/N_F}
\sim \Biggl (\frac{\mh}{\la} \Biggr )^{\bd/3N_c}\ll 1\,,\quad
{\hat z}_M={\hat z}_M(\la,\,\ogh)=1/{{\hat z}_q}^2\gg 1\,.
\eeq
Therefore,
\bq
\frac{\mu(M_{\it hl})}{\la}\sim \frac{\mu(N_{\it lh})}{\la}\sim \exp \{-\nd/14\bd \}
\,\Biggl (\,\frac{\mh}{\la}\Biggr )^{N_F/3N_c}\sim\frac{\it \ogh}{\la}\,.\label{(5.11)}
\eq

3)\,\,\,Because the ${\ov q}_{\it h}$ and  $q^{\it h}$  quarks are higgsed,  $N_{\it h}^2$ pseudo-Goldstone mesons $N_{\it hh}$ (nions) appear. After integrating out the heavy gluons and their superpartners, the Lagrangian of remained degrees of freedom takes the form
\bbq
K\simeq {\hat z}_M {\rm Tr}\Biggl (\frac{M^{\dagger}_{\it hh}M_{\it hh}+ M^{\dagger}_{\it ll}M_{\it ll}}{Z^2_q\la^2}\Biggr )+\,
{\hat z}_q\,2\,{\rm Tr}\sqrt {N^{\dagger}_{\it hh} N_{\it hh}}\,+{\hat z}_q {\rm Tr\,}_{\it l}\,\Biggl ( q^{\dagger} q + {\ov q}^
{\dagger} {\ov q}\Biggr )\,,
\eeq
\bq
\w= -\frac{2\pi}{\ov \alpha^{\,\prime}(\mu)}{\ov S}^{\,\prime}+{\rm Tr}\,\Biggl (\frac{M_{\it hh} N_{\it hh}}{Z_q\la}\Biggr )+{\rm Tr\,}_{\it l}
\Biggl (\frac{{\ov q}\, M_{\it ll}\, q} {Z_q\la}\Biggr )+\,{\rm Tr} \,\Biggl ( \ml M_{\it ll}+\mh M_{\it hh}\Biggr )\,, \label{(5.12)}
\eq
where ${\ov S}^{\,\prime}$ includes the field strengths of the remaining $SU( \nd^{\,\prime})$ dual gluons
with $\nd^{\,\prime}=\nd-\nh=\nl-N_c$ dual colors, and $q^l,\, {\ov q}_{\it \ov l}$ are the still active $l$ - flavored dual quarks
with unhiggsed colors and, finally, the nions $N_{\it hh}$ are "sitting down" inside ${\ov \alpha}^{\,\prime}(\mu)$.

At lower scales $\mu<\ogh$, the mions $M_{\it hh}$ and nions $N_{\it hh}$ are frozen and don't evolve, while the gauge coupling decreases logarithmically in the interval $\olp <\mu<\ogh$. The numerical value of the pole mass of ${\ov q}_{\it \ov l}\,,\, q^{\it l}$ - quarks is
\bq
\frac{\olp}{\la}=\frac{\langle M_{\it ll}\rangle}{Z_q\la^2}\,\frac{1}{{\hat z}_{q}\,z^{\,\prime\prime}_{q}}\sim\exp \Bigl \{\,\frac{\nd}{7\bd}\,\Bigr \}\,\Biggl [\, \Bigl(r\Bigr )^{\frac{\nl-N_c}{N_c}}\,\Biggl (\frac{\mh}{\la} \Biggr )^{N_F/3N_c}\,\Biggr ]/z_q^{\,\prime\prime}\,, \label{(5.13)}
\eq
\bbq
\olp\sim \exp \Bigl \{\,\frac{\nd}{7\bd}\Bigr \}\,\mulp\gg \mulp\,,\quad\quad z^{\,\prime\prime}_{q}=z^{\,\prime}_{q}(\ogh\,,{\olp})\simeq
z^{\,\prime}_{q}(\mhp\,,\mulp)=z^{\,\prime}_{q}\,,
\eeq
where $z^{\,\prime\prime}_{q}\ll 1$ is the logarithmic renormalization factor of $\dl\,,\odl$ quarks.

After integrating out the quarks ${\ov q}_{\it \ov l}\,,\, q^{\it l}$ as heavy ones at $\mu<\olp$, the dual $SU(\nl-N_c)$ Yang-Mills theory remains with the scale factor of its gauge coupling $\langle -{\ov \Lambda}_L\rangle=\lym$ (and with nions $N_{\it hh}$ and mions $M_{\it ll}$ "sitting down" on ${\ov \Lambda}_L$). Finally, integrating out the dual gluons via the VY procedure, we obtain (\,recalling that all fields in \eqref{(5.12)} and \eqref{(5.14)} are normalized at $\mu=\la$, and $\lym/\la=(r)^{\nl/3N_c}\,(\mh/\la)^{N_F/3N_c}\,,\,\,\langle N_{\it hh} \rangle=-Z_q\mh\la $\,)
\bq
K={\hat z}_M {\rm Tr}\Biggl (\frac{M^{\dagger}_{\it hh}M_{\it hh}+z_{M}^{\,\prime\prime} M^{\dagger}_{\it ll}M_{\it ll}}{Z^2_q\la^2}\Biggr )+\,
{\hat z}_q\,2\,{\rm Tr}\sqrt {N^{\dagger}_{\it hh} N_{\it hh}}\,,\label{(5.14)}
\eq
\bbq
\w={\rm Tr}\,\Biggl (\frac {M_{\it hh}N_{\it hh}}{Z_q\la}\Biggr ) -(\nl-N_c)\,\lym^3\Biggl (\det \frac{\langle N_{\it hh}\rangle}{N_{\it hh}}\det \frac{M_{\it ll}}{\langle M_{\it ll}\rangle} \Biggr )^{1/(\nl-N_c)}+\,{\rm Tr} \,\Biggl ( \ml M_{\it ll}+\mh M_{\it hh}\Biggr )\,,
\eeq
where $z^{\,\prime\prime}_{M}=z^{\,\prime\prime}_{M}(\ogh\,,{\olp})\simeq z^{\,\prime}_{M}( \mgh\,,
{\mulp})=z^{\,\prime}_M\gg 1$ is the logarithmic renormalization factor of $M_{\it ll}$ mions.
\footnote{\,
The second term of the superpotential in \eqref{(5.14)} can equivalently be written as\,:
\bbq
-(\nl-N_c)\,\Biggl ( \frac{\det {M_{\it ll}}}{\la^{\bo}\,{\det \Bigl (-N_{\it hh}/Z_q\la\Bigr )}}\Biggr )^{1/(\nl-N_c)}
\eeq
}

The masses obtained from \eqref{(5.14)} look then as follows\,:
\bq
\frac{\mu(M_{\it hh})}{\la}\sim \frac{\mu(N_{\it hh})}{\la}\sim \Biggl ({\hat z}_q\frac{|\langle N_{\it hh}\rangle|}{\la^2} \Biggr )^{1/2}\sim \exp \Bigl \{-\nd/14\bd \Bigr \}\,\Biggl (\,\frac{\mh}{\la}\Biggr )^{N_F/3N_c}\sim\frac{\it\ogh}{\la}\,,\label{(5.15)}
\eq
\bbq
\frac{\mu(M_{\it ll})}{\la}\sim \frac{Z_q^2}{z_{M}^{\,\prime\prime}}\frac{\ml\,\la}{{\hat z}_{M}\langle M_{\it ll}\rangle}\sim \exp \Bigl \{-\frac{2\nd}{7\bd} \Bigr \}\Bigl (r \Bigr )^{\frac{2N_c-\nl}{N_c}}\Biggl (\frac{\mh}{\la} \Biggr )^{N_F/3N_c}/z^{\,\prime\prime}_M\,.
\eeq
\vspace{0.5cm}

On the whole for Section 5.2, when we started directly from the dual theory with $\nd=(N_F-N_c)$
colors, $N_F=(\nl+\nh)$ quarks $\ov q\,,q$ and  $N_F^2$ mions $M$, its mass spectrum is as follows.\\

{\bf 1)\,} The sector of heavy masses includes\,:\, a) $(2\nd\nh-\nh^2)$ of massive dual gluons and
their scalar superpartners;\, b)\, $2\nl\nh$ hybrid scalar mions $M_{\it hl}+M_{\it lh}$,\, c)\, $2\nl\nh$ hybrid
scalar nions $N_{\it lh}+ N_{\it hl}$ (these are $q^{\it l}\,,\ov q_{\it \ov l}$\,\,
quarks with higgsed colors),\, d)\, $\nh^2$ scalar mions $M_{\it hh}$ and $\nh^2$ of scalar
nions $N_{\it hh}$. All these particles, with {\it specific numbers of each type, definite
spins, and other quantum numbers}, have definite masses of the same order
\bbq
\sim \ogh\sim \exp\{-\nd/14\bd\}(\mh/\la)^{N_F/3N_c}\la\sim\exp \{-\nd/14\bd \}\,\mhp\ll\mhp\sim(1/r)^{\nl/3N_c}\lym\,.
\eeq

All these particles are {\it interacting only weakly}\,, with both their gauge and Yukawa couplings

${\ov a}_*\sim a_{f}^{*}\sim \bd/N_F\ll 1$.\\

{\bf 1')}\,\,Because the dual quarks $q^{\it h}\,,{\ov q}_{\it h}$ are higgsed, one can imagine
that solitonic excitations also appear in the form of monopoles of the dual gauge group (its
broken part), see e.g. section 3 in \cite{ch3} and footnote 6 therein.
These dual monopoles are then confined and can, in principle, form a number of additional hadrons
$H_{\it h}^{\,\prime}$. Because the dual theory is weakly coupled at the scale of higgsing,
$\mu\sim \ogh \sim \exp \{-\nd/14\bd \}\, \mhp$, the masses of these monopoles, as well as the
tension $\sqrt {\,\ov \sigma}$ of strings confining them (with our exponential accuracy in parametrical
dependence on $\nd/\bd\gg 1$), are also $\sim \ogh$\,. Therefore, the mass scale of
these hadrons $H_{\it h}^{\,\prime}$ is also $\sim \ogh\ll \mhp$\,. We even assume (in
favor of the duality) that, with respect to their quantum numbers, these hadrons $H_{\it h}^
{\,\prime}$ can be identified in some way with the direct hadrons made of the quarks ${\ov Q}^{\it h}\,,
\qh$. But even then, the masses of $H_{\it h}^{\,\prime}$ are {\it parametrically smaller}
than those of various direct hadrons made of the ${\ov Q}^{\it h}\,,\,\qh$ quarks\,,
$\mu(H^{\,\prime}_{\it h})\sim \exp \{-\nd/14\bd\}\,\mhp\ll \mhp$.

Besides, because the chiral symmetry of $\,{\it \ov l}\,,\,\it l$\, flavors and the $R$-charges of the
lower energy theory at $\olp<\mu < \ogh$ {\it remain unbroken} and the $\it l$ - flavored dual quarks
are not higgsed and remain effectively massless in this interval of scales\,, no possibility is seen in this
dual theory in Section 5.2 for the appearance of $\it{\ov l}\,, l$\,-flavored chiral hadrons
$H^{\,\prime}_{\it l}$ (mesons and baryons) with the heavy masses $\sim \ogh$ that could be identified
with a large number of various direct $\it {\ov l}\,, l$\, - flavored chiral hadrons (mesons and baryons)
made of ${\ov Q}^{\it \ov l}\,, \ql$ quarks (and $W_{\alpha}$) that appear when the direct theory is dualized
in variant "b" (not even speaking about their parametrically different mass scales, $\ogh\ll \mhp$\,,
see point {\bf 1'} in Section 5.1). This shows the self-contradictory character of duality in variant "b" =
"confinement without chiral symmetry breaking".\\

All other particles in the mass spectrum of the dual theory in this section 5.2 constitute the sector of lighter
particles, with their masses being parametrically smaller than $\ogh$.\\

{\bf 2)} The next mass scale is formed by a large number of ${\it l}$\,-flavored dual mesons and ${\rm b}_{\it l}
\,,{\rm \ov b}_{\it l}$ baryons made of non-relativistic (and weakly confined, the string tension
$\sqrt \sigma\sim \lym\ll \olp$)\, dual ${q^{\it l}}\,,\ov q_{\it \ov l}$ quarks with $\nd^{\,\prime}=(\nl-N_c)$
unhiggsed colors. The pole masses of these $q^{\it l}\,,\ov q_{\it \ov l}$ quarks are
$$\olp \sim \exp\{3\nd/14\bd \}\,(r)^{(\nl-N_c)/N_c}\,{\it\ogh}/z^{\,\prime}_{q}\ll {\it\ogh}$$
(at $r\ll r_l$, see \eqref{(5.16)} below).\\

{\bf 3)} Next, there is a large number of gluonia with the mass scale $\sim \lym\ll\olp$.\\

{\bf 4)} Finally, the lightest are $\nl^2$  scalar mions $M_{\it ll}$ with masses
$$\mu(M_{\it ll})\sim \exp \{-2\nd/7\bd \} (r )^{2\Delta}\,\lym/z^{\,\prime}_M\ll \lym\,,\quad 0<\Delta=\frac{3N_c-2\nl}{3N_c}<\frac{1}{3}\,.$$
\vspace{3mm}

Comparing the mass spectra of two supposedly equivalent descriptions in Sections 5.1  and 5.2
above, it is seen that the masses are clearly different parametrically, in powers of the parameter
$Z_q\sim\exp\{-\nd/7\bd\}\ll 1$. Besides, the theory described in Section 5.2 contains very specific definite numbers of fields with fixed quantum numbers and spins and with the definite masses $\sim\ogh\sim Z_q^{1/2}
\mhp\ll\mhp$, all {\it parametrically weakly interacting} (see point {\bf 1} in Section 5.2). No analog of
these distinguished particles is seen in Section 5.1. Instead, there is a large number of ${\it h}$ - flavored hadrons with the masses $\sim\mhp$ and with various spins, all strongly interacting with the coupling $a(\mu\sim \mhp)\sim 1$.

Finally, and we consider this to be of special importance,
{\it there are no heavy ${\it \ov l}\,, l$ -flavored hadrons} $H^{\,\prime}_{\it l}$ in the theory described
in Section 5.2, which have the appropriate conserved (in the interval of scales $\olp<\mu < \ogh$)
chiral flavors ${\it \ov l}\,,{\it l}$  and $R$ - charges, such that they can be associated with a
large number of various heavy flavored chiral hadrons made of ${\ov Q}^{\it \ov l}\,, Q_{\it l}$ -quarks (and
$W_{\alpha}$), which are present in Section 5.1 dualized in variant "b". This shows that the duality
in the variant "b" = "confinement without chiral symmetry breaking" is not self-consistent.
\footnote{\,
It is not difficult to see that a similar situation with the ${\it {\ov l}\,,l}$\,-flavored chiral hadrons that can be made of the ${\ov Q}^{\it \ov l}
\,, Q_{\it l}$ - quarks (and $W_{\alpha}$\,) occurs in variant "b" also in the scenario $\#1$ considered in \cite{ch1, ch2} (see section 3b in \cite{ch2}
for a similar regime). The analog of Section 5.1 therein is exactly the same as in this paper. The analog of Section 5.2 is different, because the {\it h} -flavors are not higgsed, but instead form the diquark condensate; qualitatively, however, the situation is the same, and the difference between analogs of Section 5.1 and 5.2 is only more prominent in scenario $\#1$. \label{(f7)}
}
This agrees with some general arguments presented in \cite{ch1} (see Section 7 therein, it is also worth
recalling that {\it these arguments were not related with the use of scenario $\#1$ with the diquark
condensate}) that the duality in variant "b" cannot be realized because, in the theory
{\it with unbroken chiral flavor symmetries and R-charges} and effectively massless quarks,
the masses of flavored and R - charged chiral hadron superfields with various spins cannot be made "of nothing".\\

This is not the whole story, however. For the mass $\ogh$ of gluons in the dual theory to be the largest physical mass $\mu_H$ as was used in Section
5.2 above, the parameter $r=\ml/\mh$ has to be taken sufficiently small (see \eqref{(5.9)},\eqref{(5.13)}; from now on, the non-leading
effects due to logarithmic factors like $z^{\,\prime}_q\simeq z^{\,\prime\prime}_q$  are ignored)\,:
\bq
\frac{\olp}{\ogh}\ll 1 \, \ra \, r \ll r_{\it l}=\Biggl (\,z^{\,\prime}_q\,
\exp \Bigl \{-\frac{3\nd}{14\bd}\Bigr \}\, \Biggr )^{\frac{N_c}{\nl-N_c}}\sim
\exp \Biggl \{-\frac{3\nd}{14\bd}\,\frac{N_c}{\nl-N_c}\,\Biggr \}\ll 1\,. \label{(5.16)}
\eq
We trace below the behavior of the direct and dual theories in the whole interval $r_{\it l}<r<1$.
\footnote{\,
For this, it is convenient to keep $\mh$ intact while $\ml$ will be decreased, starting from $\ml=\mh$.
}

As regards the direct theory (see Section 5.1 above), its regime and all hierarchies in the mass
spectrum remain the same for any value of $r<1$\,, i.e. the pole mass $\mhp$ of the $\qh\,,\oqh$
quarks becomes the largest physical mass $\mu_H$ already at $r<1/(\rm several)$, and so on.

 But this is not the case for the dual theory (see Section 5.2 above). At $r_{\it l}<r<1/
(\rm several)$\,, the pole mass $\hlp$ of $\dl\,,\odl$ quarks remains the largest physical mass
\bq
\frac{{\hlp}}{\la}\sim\frac{\langle M_{\it ll}\rangle}{Z_q\la^2}\,\Biggl (\frac{\la}{\hlp}\Biggr )^{\bd/N_F}\sim\exp \Biggl \{\,\frac{\nd}{7\bd}\,\Biggr \}\,
\Bigl( r \Bigr )^{\Bigl (\frac{\nl-N_c}{N_c}\frac{N_F}{3\nd}\Bigr )}\,\Biggl (\frac{\mh}{\la} \Biggr )^{N_F/3N_c}\sim \frac{\olp}{\la}\,.\label{(5.17)}
\eq

Already this is sufficient to see a qualitative difference between the direct and
dual theories. The {\it hh}\,-flavored hadrons in the direct theory have the largest masses, while
the {\it ll}\,- flavored hadrons are the heaviest ones in the dual theory.

We make now some rough additional estimates in Section 5.2 at $r>r_{\it l}$. After integrating out the
heaviest quarks $\dl\,,\odl$ at the scale $\mu\sim \hlp$, the dual theory remains with
$\nd$ colors and $\nh<\nd$  dual quarks ${\ov q}_{\it \ov h}\,,\, q^{\it h}$ (and mions $M$),
and with $\bd^{\prime\prime}=(3\nd-\nh)>0$. It is in the weak-coupling logarithmic regime
at $\mu^{\,\prime}_H<\mu<\hlp$\,,\,\,\, where $\mu^{\,\prime}_H$ is the highest mass scale in the remaining
theory. The new scale factor $\Lambda^{\,\prime}_q$ of its gauge coupling can be found from
\bq
\bd^{\,\prime\prime}\ln\Biggl (\frac{\hlp}{\Lambda^{\,\prime}_q} \Biggr )\simeq \frac{2\pi}{\ov\alpha_*}=\frac{3\nd^2}{7\bd}\quad \ra \quad \frac{\Lambda^{\,\prime}_q}{\hlp}\sim\exp\Biggl \{-\frac{3\nd^2}{7\bd\bd^{\,\prime\prime}}\Biggr \}\ll 1\,. \label{(5.18)}
\eq

If (see below) $r_{\it l}\ll r_{\it h}<r<1/(\rm several)$, then $\mu^{\,\prime}_H=\hhp>
\Lambda^{\,\prime}_q$, where $\hhp$ is the pole mass of $\hd\,,\ohd$ quarks which are
in the HQ-phase (i.e. not yet higgsed ). Roughly, $\hhp\sim r\,\hlp$, whence
\bq
\frac{\hhp}{\Lambda^{\,\prime}_q}>1  \quad \ra \quad r>r_{\it h}\sim \exp \Biggl \{-\frac
{3\nd^2}{7\bd\bd^{\,\prime\prime}} \Biggr \}\gg r_{\it l}\,. \label{(5.19)}
\eq

In the interval $r_{\it h}<r<1/(\rm several)$, the mass spectrum of the dual theory is qualitatively
not much different from the case $r=1$ (see Section 4). The heaviest are the {\it ll} - hadrons,
then the {\it hl} - hadrons, then the {\it hh} - hadrons, then gluonia and, in addition, there are the
mions $M_{\it hh}\,,\,M_{\it hl}\,,\, M_{\it ll}$ with their masses
$$\mu(M_{\it hh})\sim\mh^2/\mu_{o}\,,\,\,\,\mu(M_{\it hl})\sim \mh \ml/\mu_{o}\,,\,\,\,\mu(M_{\it ll})
\sim \ml^2/\mu_{o}\,,$$
where
$$\mu_{o}= z_{M}(\la,\hlp)\lym^{3}/Z_q^2\la^2\,,\quad z_{M}(\la,\hlp)\sim (\la^2/\mh\ml)^{\bd/3N_c}\gg 1.$$

As $r$ decreases further, a phase transition occurs from the $\rm HQ_{\it h}$ phase to the
$\rm Higgs_{\it h}$ phase at $r\sim r_{\it h}$\,, i.e. after $\hhp >\Lambda^{\prime}_q$
becomes $\hhp<\Lambda^{\,\prime}_q$ and the quarks $\hd\,,\ohd$ are higgsed.~
\footnote{\,
As was argued in \cite{ch3} (see section 3 therein), the transition proceeds through formation
of a mixed phase in the threshold region $\Lambda^{\,\prime}_q/(\rm several)<\hhp <(\rm several)
\Lambda^{\,\prime}_q$\, (i.e. ${\it r}_{\it h}/(\rm several)<{\it r}<(\rm several)
{\it r}_{\it h}$\, here)\,.
}
But even then the $\dl\,,\odl$ quarks remain the heaviest ones. And only when $r$ becomes $r
<r_{\it l}\ll r_{\it h}\ll 1$, the gluon mass $\ogh$ becomes the largest and the mass
spectrum of the dual theory becomes that described above in this section 5.2.\\

\section{ Direct theory. Unequal quark masses,\\ {\boldmath  $3N_c/2<N_F<3N_c\,,\,\,\bo/N_F=O(1)\,,\,\,\nl>N_c$}}

This section continues the preceding one, but we forget about any dualizations in what follows and
we deal with the direct theory as it is, i.e. in variant "a" (see \cite{ch1}, section 7). This means
that after the heavy quarks $\qh\,,\oqh$ have been integrated out at $\mu<\mhp$, all particle
masses in the lower energy theory with $N_c$ colors and $N_c<\nl<3N_c/2$ lighter $\ql\,,
\oql$ quarks are parametrically smaller than the scale $\la^{\,\prime}\sim \mhp$ (see \eqref{(5.1)}
above), and at $\mu<\la^{\,\prime}$ this lower energy theory enters the strong-coupling regime
with $a(\mu\ll\la^{\,\prime})\gg 1$.
\footnote{\,
Hereafter, to have definite answers, we use the anomalous quark dimension $1+\gamma_Q(N_F\,,N_c\,,a(\mu)\gg 1)
=N_c/(N_F-N_c)$, and the strong coupling $a(\mu)\gg 1$ given in eq.(7.4) in \cite{ch1}.
}

This lower-energy theory is in the HQ-phase in the variant "a" (see footnote \ref{(f4)}), and the pole mass of
$\ql\,,\oql$ quarks is
\bbq
\frac{\mlp}{\la^{\,\prime}}\sim\Biggl (\frac{\ml(\mu=\la^{\,\prime})}{\la^{\,\prime}}=r\,\Biggr )
\Biggl (\frac{\la^{\,\prime}}{\mlp}\Biggr )^{\gamma^{\,\prime}_Q}\quad \ra \quad \frac{\mlp}
{\la^{\,\prime}}\sim\Bigl (r \Bigr )^{(\nl-N_c)/N_c}\ll 1\,,\quad r\ll 1\,,
\eeq
\bq
{\rm b}^{\,\prime}_o=3N_c-\nl,\quad  1+\gamma^{\,\prime}_Q=\frac{N_c}{\nl-N_c}\,, \quad \nu=\frac{N_F
\gamma_Q^{\,\prime}-{\rm b}^{\,\prime}_{\rm o}}{N_c}=\frac{3N_c-2\nl}{\nl-N_c}\,. \label{(6.1)}
\eq

The coupling $a_{+}(\mu=\mlp)$ is large \cite{ch1}
\bq
a_{+}(\mu=\mlp)\sim\Biggl (\frac{\la^{\prime}}{\mu=\mlp} \Biggr )^{\nu}\sim\Bigl (\,\frac{1}{r}\,
\Bigr )^{(3N_c-2\nl)/N_c}\gg 1\,.\label{(6.2)}
\eq

Hence, after integrating out all $\ql\,,\oql$ quarks as heavy ones at $\mu<\mlp$, we are left
with the pure $SU(N_c)$ Yang-Mills theory, but it is now in the {\it strong-coupling regime}.
This is somewhat unusual, but there is no contradiction because this perturbative strong-coupling
regime with $a_{YM}(\mu)\gg 1$ is realized in a restricted interval of scales only, \,$\lym\ll
\mu<\mlp\ll \la^{\prime}$. It follows from the NSVZ $\beta$ -function \cite{NSVZ} that the
coupling $a_{YM}(\mu\gg \lym)\gg 1$ is then given by
\bbq
a_{YM}(\mu=\mlp)=\Biggl (\frac{\mu=\mlp}{\lambda_{YM}} \Biggr )^3=
a_{+}(\mu=\mlp)\,\ra \, \frac{\lambda_{YM}}{\la}=\frac{\lym}{\la}=\Biggl [ \Bigl
(\frac{\ml}{\la}\Bigr )^{\nl}\Bigl (\frac{\mh}{\la}\Bigr )^{\nh}\Biggr ]^{\frac{1}{3N_c}},
\eeq
\bq
\frac{\lym}{\mlp}=r^{\frac{3N_c-2\nl}{3N_c}}\ll 1,\quad a_{YM}(\lym\ll\mu<\mlp)=a_{YM}(\mu=\mlp)
\Biggl (\frac{\mu}{\mlp}\Biggr )^3=\Biggl (\frac{\mu}{\lym} \Biggr )^3 \label{(6.3)}
\eq
and it now {\it decreases} from $a_{YM}(\mu=\mlp)\gg 1$ to $a_{YM}(\mu\sim \lym)\sim 1$
as $\mu$ decreases, after which the non-perturbative effects become essential.

Therefore, decreasing the scale $\mu$ from $\mu=\mlp$ to $\mu <\lym$, integrating out all gauge
degrees of freedom except for the one whole field $S\sim W_{\alpha}^2$\,, and using
the VY-form of the superpotential of the field $S$ \cite{VY}, we obtain the standard
gluino condensate, $\langle S\rangle=\lym^3$.

To verify the self-consistency, we have to estimate the scale $\mu_{\rm gl}$ of the
possible higgsing of $\ql\,,\oql$ quarks. This estimate looks as
\bq
\mu_{\rm gl}^2\sim a_{+}(\mu=\mu_{\rm gl})\langle \oql \ql \rangle_{\mu=\mu_{\rm gl}}\sim \Bigl (\mlp \Bigr )^2\,,\quad a_{+}(\mu=
\mu_{\rm gl})\sim\Biggl ( \frac{\la^{\,\prime}\sim\mhp}{\mu_{\rm gl}}\Biggr )^{\nu}\,, \label{(6.4)}
\eq
\bbq
\langle \oql \ql \rangle_{\mu=\mu_{\rm gl}}\sim\langle \oql \ql \rangle_{\mu=\la}\,\Biggl(\frac{
\la^{\,\prime}}{\la} \Biggr )^{\gamma_Q=(\bo/N_F)}\,\Biggl (\frac{\mu_{\rm gl}}{\la^{\,\prime}} \Biggr )^
{\gamma^{\,\prime}_Q}\,,\quad \langle \oql \ql \rangle_{\mu=\la}=\frac{\lym^3}{\ml}\,.
\eeq

As before, we assume that really $\mu_{\rm gl}=\mlp/(\rm several)$, and hence the $\ql\,,\oql$
quarks are not higgsed and the $HQ_{\it l}$ phase is self-consistent, see footnote \ref{(f4)}.\\

On the whole, all quarks of the direct theory are in the HQ phase (and are therefore confined, but the string tension
is small in comparison with quark masses, $\sqrt\sigma\sim \lym\ll\mlp\ll\mhp$). The mass spectrum then includes
\,:\, 1)\, a large number of heavy $\it hh$\,-flavored mesons with the mass scale $\sim \mhp\ll
\la$\,;\,\, 2)\, a large number of hybrid $\it hl$\,-mesons and baryons $B_{\it hl}\,,
{\ov B}_{\it hl}$ with the same mass scale $\sim \mhp$\,;\,\, 3)\, a large number of
$\it ll$\,-flavored mesons and baryons $B_{\it l}\,,{\ov B}_{\it l}$ with the mass scale $\sim\mlp\ll\mhp$\,,
and finally,\,\, 4)\, gluonia, which are the lightest, with their mass scale $\sim \lym \ll \mlp$.

\section{Direct and dual theories. Equal quark masses, \\ {\hspace*{3cm} \boldmath $N_c<N_F<3N_c/2$}}

As regards the direct theory, this case is obtained from the one in the preceding section by a simple change of notations.
The quark pole masses are
\bq
\frac{\mq}{\la}\sim\frac{m_Q}{\la}\Biggl (\frac{\la}{m_Q=m_Q(\mu=\la)}\Biggr )^{\gamma_Q}\sim\Biggl (\frac{m_Q}{\la}\Biggr )^
{\frac{1}{1+\gamma_Q}=\frac{\nd}{N_c}},
\label{(7.1)}
\eq
while the gluon masses due to possible higgsing of quarks look as
\bbq
\mu_{\rm gl}^2\sim a(\mu=\mu_{\rm gl})\langle {\ov Q} Q \rangle_{\mu=\mu_{\rm gl}}\,,\quad
a(\mu=\mu_{\rm gl})\sim\Biggl (\frac{\la}{\mu_{\rm gl}}\Biggr )^{\nu}\,,\quad \nu=\frac{N_F\gamma_Q-{\rm b_o}}
{N_c}\,,
\eeq
\bq
\langle {\ov Q} Q \rangle_{\mu=\mu_{\rm gl}}= \langle {\ov Q} Q\rangle_{\mu=\la}\,\Biggl(\frac{
\mu_{\rm gl}}{\la} \Biggr )^{\gamma_Q}.\label{(7.2)}
\eq
It follows from \eqref{(7.2)} that
\bq
\frac{\mu_{\rm gl}}{\la}\sim \Biggl (\frac{m_Q}{\la}\Biggr )^{\frac{1}{1+\gamma_Q}}\sim \frac{\mq}{\la}\,,
\quad \frac{1}{1+\gamma_Q}=\frac{N_F-N_c}{N_c}\,\,\,\,{\rm for}\,\,\,\, \gamma_Q=\frac{2N_c-N_F}{N_F-N_c}\,,\label{(7.3)}
\eq
and, as previously, we assumed that $\mq=(\rm several)\,\mu_{\rm gl}$ and the quarks are not higgsed but confined
(see footnote \ref{(f4)}). After all quarks are integrated out as heavy ones at $\mu<\mq$, we are left with the $SU(N_c)$ Yang-Mills theory in the strong-coupling regime and with the scale factor $\lym\ll \mq$ of its gauge coupling, and so on.\\

As regards the dual theory, its mass spectrum has been described in \cite{ch1} (see section 7) and we only recall
it here briefly. The Lagrangian at $\mu=\la$ is taken in the form
\bbq
{\ov K}={\rm Tr}\Biggl ( q^\dagger e^{\ov V} q + {\ov q}^\dagger{\ov q}\Biggr )+\frac{{\rm Tr}\Bigl ( M^{\dagger} M\Bigr )}{\la^2}\,,\quad
{\ov \w}= -\,\frac{2\pi}{{\ov \alpha}(\mu=\la)}\,{\ov S}+\frac{{\rm Tr}\Bigl ({\ov q}\,
M\, q \Bigr )}{\la}  + {\rm Tr} \Bigl ( m_Q\, M \Bigr )\,,
\eeq
\bq
\langle M(\mu=\la)\rangle =\cm^2\,,\quad  {\ov a}(\mu=\la)={\nd}{\ov \alpha}(\mu=\la)/2\pi=O(1)\,.\label{(7.4)}
\eq

The dual quarks are in the HQ  phase and are therefore confined, and their pole masses are (from now on and in this
section we neglect the logarithmic renormalization factors $z_q$ and $z_M$ for simplicity)
\bq
\frac{\mu^{\rm pole}_q}{\la}=\frac{\langle M(\mu=\la)\rangle =\cm^2=\langle {\ov Q}Q(\mu=\la)\rangle}{z_q(\la,
\,\mu^{\rm pole}_q)\la^2}\sim \Bigl (\frac{m_Q}{\la}\Bigr )^{\frac{N_F-N_c}{N_c}}.\label{(7.5)}
\eq

After integrating out the dual quarks as heavy ones, we are left with the dual gauge theory with $\nd$ colors and with the scale factor $\langle\Lambda_L(M)\rangle=\lym$ of its gauge coupling, and with the frozen mion fields $M$. Finally, after integrating out the dual gluons by means of the VY- procedure \cite{VY}, we obtain the Lagrangian of mions
\bq
{\ov K}=\frac{1}{\la^2}\,{\rm Tr\,\Bigl (M^{\dagger}M\Bigr )}\,,\quad
{\ov \w}= -\nd \, \Biggl ( \frac {\rm \det\, M}{\la^{\bo}} \Biggr )^{1/\nd}
+m_Q{\rm Tr  \,M}\,,\quad \mu \ll \Lambda_{YM}\,\,.\label{(7.6)}
\eq
It describes the mions ${\rm M}$ with the masses
\bq
\mu_M\sim m_Q\Biggl (\frac{\la^2}{\cm^2}\Biggr )\sim m_Q \Biggl (\frac{\la}{m_Q}\Biggr )^
{\frac{N_F-N_c}{N_c}}\,,\quad m_Q\ll \mu_M \ll \Lambda_{YM}\,.\label{(7.7)}
\eq
Clearly, there is no analog of these parametrically light particles in the direct theory.

\section{ Direct and dual theories. Unequal quark masses. \\ {\hspace*{3cm}\boldmath $N_c<N_F<3N_c/2\,,\,\,\nl>N_c$}}

As regards the direct theory, the mass spectrum in this case with $r=\ml/\mh< 1$ is not much different from the
one in the preceding section. All quarks are in the HQ  phase, and the highest physical mass is the pole mass of
{\it h} - quarks
\bq
\mhp=\mh\Biggl (\frac{\la}{\mhp} \Biggr )^{\gamma_{+}}\quad \ra \quad \frac{\mhp}{\la}=
\Biggl (\frac{\mh}{\la}\Biggr )^{(N_F-N_c)/N_c}\,,\quad \gamma_{+}=\frac{(2N_c-N_F)}{(N_F-N_c)}\,.\label{(8.1)}
\eq

After integrating out the $\qh\,,\oqh$  quarks as heavy ones at $\mu<\mhp$, the lower energy theory remains with $N_c$ colors and $N_c<\nl<3N_c/2$
\, {\it l} - quarks, in the strong coupling regime with $a(\mlp<\mu<\mhp)\gg 1$, see sections 6, 7. The next independent physical scale is the pole mass
of {\it l} - quarks\, ($\gamma_{-}=(2N_c-\nl)/(\nl-N_c)$\,)
\bq
\mlp\sim\Biggl (\ml(\mu=\mhp)=r\,\mhp \Biggr )\,\Biggl (\frac{\mhp}{\mlp} \Biggr )^{\gamma_{-}} \quad \ra \quad\Bigl (\frac{\mlp}{\mhp}\Bigr )
\sim\Bigl ( r\Bigr )^{\frac{\nl-N_c}{N_c}}\ll 1\,. \label{(8.2)}
\eq

After integrating out the {\it l} - quarks as heavy ones at $\mu<\mlp$, leaves us with the $SU(N_c)$ YM  theory in
the strong-coupling regime with the large gauge coupling. The scale factor $\lym^{\,\prime}$ of its gauge coupling is determined from (see sections 6, 7)
\bq
\Biggl (\frac{\mlp}{\lym^{\,\prime}}\Biggr )^3=\Biggl (\frac{\la}{\mhp} \Biggr )^{\nu_{+}}
\Biggl (\frac{\mhp}{\mlp} \Biggr )^{\nu_{-}} \quad \ra \quad \lym^{\,\prime}=\lym=\Biggl (\la^
{\bo}\ml^{\nl}\mh^{\nh}\Biggr )^{1/3N_c}\,.\label{(8.3)}
\eq
\bbq
\nu_{+}=\frac{3N_c-2N_F}{N_F-N_c}\,,\quad \nu_{-}=\frac{3N_c-2\nl}{\nl-N_c}>\nu_{+}\,\,.
\eeq

To verify the self-consistency, we also estimate the gluon masses due to the possible higgsing
of the $\qh$ and/or $\ql$ - quarks. As for the $\qh$ - quarks\,,
\bq
\frac{\mgh^2}{\la^2}\sim \Biggl [\,a_{+}(\mu=\mgh)=\Biggl (\frac{\la}{\mgh} \Biggr )^{\nu_{+}}
\, \Biggr ]\,\frac{\langle{\ov Q}^{\it h}\qh \rangle}{\la^2}\,\Biggl (\frac{\mgh}{\la}\Biggr )
^{\gamma_{+}}\, \ra \,\, \frac{\mgh}{\mhp}\sim \Bigl (\, r \,\Bigr )^{\nl/N_c}\ll 1\,,\label{(8.4)}
\eq
and hence there is no problem, but for the $\ql$ - quarks, we now have
\bq
\frac{\mgl^2}{\la^2}\sim \Biggl [\,a_{-}(\mu=\mgl)=\Biggl (\frac{\la}{\mhp} \Biggr )^{\nu_{+}}
\Biggl (\frac{\mhp}{\mgl} \Biggr )^{\nu_{-}}\,\Biggr ]\,
\frac{\langle{\ov Q}^{\it l}\ql \rangle}{\la^2}\,\Biggl (\frac{\mhp}{\la}\Biggr )
^{\gamma_{+}}\Biggl ( \frac{\mgl}{\mhp}\Biggr )^{\gamma_{-}}\,\label{(8.5)}
\eq
\bq
\ra \quad \frac{\mgl}{\la} \sim \frac{\langle{\ov Q}^{\it l}\ql \rangle}{\la^2}\,
\sim\,\frac{\mlp}{\la}\sim\Bigl ( r\Bigr )^{\frac{\nl-N_c}{N_c}}\,\,\frac{\mhp}{\la}\,.\label{(8.6)}
\eq
As before, we assume that it is in favor of $\mlp$, i.e. $\mlp=(\rm several)\,\mgl$ (see the footnote\ref{(f4)} ).\\

On the whole in the direct theory, all quarks are in the HQ  phase and the mass spectrum consists of\,:\,
a)\, a large number of {\it hh} and hybrid {\it hl} - mesons and baryons with the mass scale $\sim \mhp
\sim \la\,(\mh/\la)^{(N_F-N_c)/N_c}\ll \la$\,,\,\, b)\, a large number of {\it ll} - mesons and baryons with the mass scale $\sim \mlp\sim (r)^{(\nl-N_c)/N_c}\,\mhp\ll \mhp$\,;\, all quarks are weakly confined, i.e. the string
tension $\sqrt\sigma\sim\lym$ is much smaller than their masses\,,\,\, c)\, gluonia, which are the lightest, with
the masses $\sim \lym \ll \mlp$.\\

In the dual theory with the Lagrangian \eqref{(7.4)}, there are several regimes depending on the value of $r=\ml/\mh< 1$.\\

i) at $r_1=(\mh/\la)^{(3N_c-2N_F)/2\nl}<r<1$,\, the hierarchy of masses looks as\,: $\mu_{q,\,\it l}>\mu_{q,\,\it h}>\mu_{\rm gl}^{\it h}$\,, where $\mu_{q,\,\it l}$ and $\mu_{q,\,\it h}$ are the masses of dual quarks and $\mu_{\rm gl}^{\it h}$ is the gluon mass due to the possible higgsing of $q_{\it h},\, {\ov q}_{\it h}$ quarks (all logarithmic renormalization effects are neglected here and below in this section for simplicity):
\bbq
\frac{\mulp}{\la}\sim \frac{\mu_{q,\,\it l}=\mu_{q,\,\it l}(\mu=\la)}{\la}=\frac{\langle{\ov Q}_{\it l}Q_{\it l}(\mu=\la) \rangle}{\la^2}\sim\Bigl (r \Bigr )^{\frac{\nl-N_c}{N_c}}\,\Bigl (\frac{\mh}{\la}\Bigr )^{\frac{N_F-N_c}{N_c}},
\eeq
\bq
\frac{\muhp}{\la}\sim \frac{\mu_{q,\,\it h}=\mu_{q,\,\it h}(\mu=\la)}{\la}\sim\frac{\langle{\ov Q}_{\it h}Q_{\it h}(\mu=\la) \rangle}{\la^2}\sim\Bigl (r \Bigr )^{\frac{\nl}{N_c}}\,\Bigl (\frac{\mh}{\la}\Bigr )^{\frac{N_F-N_c}{N_c}}\,<\, \frac{\mulp}{\la},\label{(8.7)}
\eq
\bbq
\mu_{\rm gl}^{\it h}\sim |\langle{\ov q}_{\it h}q_{\it h} \rangle|^{1/2}\sim (\mh\la)^{1/2}\,<\,\muhp\,.
\eeq
This hierarchy shows that all dual quarks are in the HQ  phase and not higgsed.\\

ii) at $r_2=(\mh/\la)^{(3N_c-2N_F)/2(\nl-N_c)}<r<r_1=(\mh/\la)^{(3N_c-2N_F)/2\nl}$, \,
the hierarchy of masses looks as\,: $\mu_{q,\,\it l}>\mu_{\rm gl}^{\it h}>\mu_{q,\,\it h}$. This shows that ${\ov q}_{\it l},\, q_{\it l}$ quarks are in the HQ  phase and are the heaviest ones, while ${\ov q}_{\it h},\, q_{\it h}$ quarks are higgsed. The phase transition of ${\ov q}_{\it h},\, q_{\it h}$ quarks from
the ${\rm HQ}_{\it h}$  phase to the ${\rm Higgs}_{\it h}$  phase occurs in the region $r\sim r_1$.\\

iii) at $r<r_2=(\mh/\la)^{(3N_c-2N_F)/2(\nl-N_c)}$,\, the hierarchy of masses looks as\,: $\mu_{\rm gl}^{\it h}>\mu_{q,\,\it l}>\mu_{q,\,\it h}$. Here, the ${\ov q}_{\it h},\, q_{\it h}$ quarks are higgsed and $\mu_{\rm gl}^{\it h}$ is the highest mass, while the ${\ov q}_{\it l},\, q_{\it l}$ quarks are lighter and are in the HQ phase. We now give some details.\\

i) After the heaviest dual $\it l$ - quarks are integrated out at $\mu<\mu_{q,\,\it l}$\,, a theory remains with $\nd$ colors and $\nh<\nd$ quarks ${\ov q}_{\it h},\, q_{\it h}$ (and mions $M$), and with the scale factor of its gauge coupling
\bq
\Bigl (\Lambda^{\,\prime}_q \Bigr )^{{\ov b}^{\,\prime}_o}=\la^{\bd}\,\mu^{\nl}_{q,\,\it l}\,, \quad {\ov b}^{\,\prime}_o=(3\nd-\nh)>0\,,\quad \Biggl (\frac{\Lambda^{\,\prime}_q}{\mu_{q,\,\it h}}\Biggr )^{{\ov b}^{\,\prime}_o/\nd}=\Bigl (\frac{r_1}{r}\Bigr )^{2\nl/N_c}<1\,.\label{(8.8)}
\eq
Hence, after integrating out the quarks ${\ov q}_{\it h},\, q_{\it h}$ as heavy ones with masses $\mu_{q,
\,\it h}>\Lambda^{\,\prime}_q$ and then gluons via the VY - procedure, we obtain the Lagrangian of mions
\bq
\ov K= \frac{1}{\la^2}\,{\rm Tr}\,\Bigl (M^{\dagger} M\Bigr )\,,\quad  {\ov \w}= -\nd \,
\Biggl ( \frac {\rm \det\, M}{\la^{\bo}} \Biggr )^{1/\nd}+{\rm Tr}\, ( m_Q M)\,\,.\label{(8.9)}
\eq
It describes mions ${\rm M}$ with the masses
\bq
\mu(M^{i}_{j})\sim \frac{m_i m_j\la^2}{\lym^3}< \lym\,\,,\quad i,\,j={\it l\,,\,h}\,.\label{(8.10)}
\eq
\vspace{3mm}

ii) The first step of integrating out the heaviest  $\it l$ - quarks with masses $\mu_{q,\,\it l}$ is the same. But now, at $r_2<r<r_1$, the next physical mass is $\ogh>\Lambda^{\,\prime}_q$ due to higgsing of dual $\it h$ - quarks, with $\nd\ra (\nd-\nh)$ and formation of $N_{\it h}^2$ nions $N_{\it  hh}$. After integrating out the heavy higgsed gluons and their superpartners with masses $\ogh\sim (\mh\la)^{1/2}$\,, and unhiggsed gluons via the VY - procedure, the Lagrangian of the remaining degrees of freedom takes the form
\bbq
{\ov K}=\frac{{\rm Tr}\,\Bigl (M^{\dagger}M\Bigr )}{\la^2}+2\,{\rm Tr}\sqrt{N^{\dagger}_{\it hh}N_{\it hh}}\,,\quad {\ov\w}= -(\nl-N_c)\,\Biggl (
\frac{\det {M_{\it ll}}}{\la^{\bo}\,{\det \Bigl (-N_{\it hh}/\la\Bigr )}}\Biggr )^{1/(\nl-N_c)}+
\eeq
\bq
+\, \frac{1}{\la}{\rm Tr}\, N_{\it hh}\Biggl ( M_{\it hh}-M_{\it hl}M^{-1}_{\it ll}M_{\it lh}\Biggr )+{\rm Tr}\Bigl
(\ml M_{\it ll}+\mh M_{\it hh} \Bigr )\,.\label{(8.11)}
\eq
From this, the masses of mions $M_{\it hh},\,M_{\it ll}$,\, hybrids $M_{\it lh},\, M_{\it hl}$, and nions $N_{\it hh}$
are
\bq
\mu(M_{\it hh})\sim \mu(N_{\it hh})\sim\Bigl (\mh \la \Bigr )^{1/2}\sim \ogh\,,\label{(8.12)}
\eq
\bbq
\mu(M_{\it lh})\sim \mu(M_{\it hl})\sim \frac{\ml\mh\la^2}{\lym^3}\,,\quad \mu(M_{\it ll})\sim \frac{\ml^2\la^2}{\lym^3}\,.
\eeq
\vspace{3mm}

iii) The heaviest particles in this region $r<r_2$ are higgsed gluons and their superpartners with masses $\ogh\sim
(\mh \la )^{1/2}$. After these have been integrated out, there remain $\nd^{\,\prime}=(\nl-N_c)$ dual colors and $\nl$
flavors with active unhiggsed colors, and the regime at $\mu<\ogh$ is IR - free logarithmic, $\bd^{\,\prime}=(3\nd^
{\,\prime}-\nl)=(2\nl-3N_c)<0$. In this theory, the next independent physical scale is given by the mass $\mu_{q,\,\it l}$
of the $\nl$ active $\it l$ - quarks with unhiggsed colors.

The masses of $2\nl\nh$ hybrid mions $M_{\it hl}+M_{\it lh}$ and $2\nl\nh$ nions $N_{\it lh}+N_{\it hl}$
(these are the dual ${\it l}$\,-quarks that have higgsed colors) are determined mainly by their
common mass term in the superpotential\,:
\bq
K_{\rm hybr}\simeq{\rm Tr}\,\Biggl (\frac{M^{\dagger}_{\it hl}M_{\it hl} }
{\la^2}\Biggr )+{\rm Tr}\,\Bigl (N^{\dagger}_{\it lh} N_{\it lh} \Bigr )\,,\quad
\w=\Bigl (\mh\la\Bigr )^{1/2}{\rm Tr}\,\Biggl (\frac {M_{\it hl}N_{\it lh}}{\la}\Biggr )\,,\label{(8.13)}
\eq
and hence their masses are
\bq
\mu(M_{\it hl})\sim \mu(M_{\it lh})\sim \mu (N_{\it hl})\sim \mu (N_{\it lh})\sim \ogh\,.\label{(8.14)}
\eq

Passing to lower scales and integrating out first the active $\it l$ - quarks as heavy ones with masses
$\mu_{q,\,\it l}$ and then the unhiggsed gluons via the VY - procedure, we obtain
\bq
K=\frac{{\rm Tr}\,\Bigl (M^{\dagger}_{\it ll} M_{\it ll}+M^{\dagger}_{\it hh} M_{\it hh}\Bigr )}{\la^2}
+2\,{\rm Tr}\sqrt{N^{\dagger}_{\it hh}N_{\it hh}}\,, \label{(8.15)}
\eq
\bbq
{\ov\w}=-(\nl-N_c)\,\Biggl ( \frac{\det {M_{\it ll}}}{\la^{\bo}\,{\det \Bigl (-N_{\it hh}/\la\Bigr )}}\Biggr )^{1/(\nl-N_c)}+\, \frac{{\rm Tr}
\,( N_{\it hh}M_{\it hh})}{\la}+{\rm Tr}\Bigl(\ml M_{\it ll}+\mh M_{\it hh}\Bigr )\,.
\eeq
From \eqref{(8.15)}, finally, the masses are given by
\bq
\mu (N_{\it hh})\sim \mu (M_{\it hh})\sim \ogh \sim \Bigl (\mh \la \Bigr )^{1/2}\,,\quad \mu(M_{\it ll})\sim \frac{\ml^2\la^2}{\lym^3}\,.\label{(8.16)}
\eq

\section{ Direct theory. Unequal quark masses. \\ {\hspace*{1cm} \boldmath $N_c<N_F<3N_c/2\,,\,\,\nl< N_c-1$}}

In this regime, at $r=\ml/\mh\ll 1$\,, the highest physical scale $\mu_{H}$ is determined by
the gluon masses $\mgl$ that arise due to higgsing of the $\ql\,,\oql$ - quarks\,:
\footnote{\,
Here and below, the value of\, $r$\, is taken to be not too small,\, such that\, $\mgl\ll \la$, i.e. $r_H\ll r\ll 1,\,
r_H=(\mh/\la)^{\rho},\, \rho=\nd/(N_c-\nl)$. At $r$ so small  that $r\ll r_H,\, \mgl\gg\la$, the quarks $\ql\,,\oql$ are higgsed in the logarithmic weak coupling region and the form of the RG flow is different, but the regime is qualitatively the same for $\mgl\ll \la$ or $\mgl\gg \la$, and nothing happens as $\mgl$ overshoots $\la$ in the scenario considered here.
}
\bbq
\frac{\mgl^2}{\la^2}\sim \Biggl [\,a_{+}(\mu=\mgl)=\Biggl (\frac{\la}{\mgl} \Biggr )^{\nu_{+}}\,\Biggr ]\,
\frac{\langle{\ov Q}^{\it l}\ql \rangle}{\la^2}\,\Biggl (\frac{\mgl}{\la}\Biggr )^{\gamma_{+}}\,\quad \ra
\eeq
\bq
\frac{\mgl}{\la} \sim \frac{\langle{\ov Q}^{\it l}\ql \rangle}{\la^2}=\frac{(\cma)^2}{\la^2}=\frac{\lym^3}
{\ml\la^2}\sim\,\Bigl (\, \frac{1}{r}\,\Bigr )^{(N_c-\nl)/N_c}\,\Biggl (\frac{\mh}{\la} \Biggr )^{(N_F-N_c)/N_c}\ll 1\,. \label{(9.1)}
\eq

The lower energy theory includes\,: $\hat N_c=(N_c-\nl)$  unbroken colors,\, $2\nl\nh$ hybrids $\Pi_{\it hl}+{\Pi}_{\it lh}$,\, $\nh$ flavors of active  $\it h$-quarks with unbroken colors and, finally,\, $\nl^2$  pions $\Pi_{\it ll},\, \langle \Pi_{\it ll}\rangle=\langle {\ov Q}^{\it l} \ql\rangle=(\cma)^2$. Their Lagrangian at $\mu<\mgl$  we write in the form (all fields are normalized at $\mu=\la$)\,:
\bbq
K=z^{+}_Q(\Pi^\dagger,\Pi)\Bigl [K_{\it ll}+K_{\rm hybr}+\dots\Bigr ]+z^+_Q(\la,\mgl)K_{\it h},\,\,\,
K_{\it h}={\rm Tr}\Bigl ({\sq}^{\dagger} {\sq} +({\sq}\ra {\oq} )\Bigr )\,,
\eeq
\bbq
K_{\rm hybr}={\rm Tr}\Biggl (\Pi^{\dagger}_{{\it l}{\it h}}\,\frac{1}{\sqrt{\Pi_{\it ll}\Pi^{\dagger}_{\it ll}}}\,\Pi_{{\it l}{\it h}}+
\Pi_{{\it h}{\it l}}\,\frac{1}{\sqrt{\Pi^{\dagger}_{\it ll}\Pi_{\it ll}}}\,\Pi^{\dagger}_{{\it h}{\it l}}\Biggr ),\quad K_{\it ll}= 2\,{\rm Tr}\sqrt{\Pi^{\dagger}_{\it ll}\Pi_{\it ll}}\,\,,
\eeq
\bq
W=-\frac{2\pi}{{\hat\alpha}(\mu)}\,\textsf{S}+W_{\Pi}+\mh {\rm Tr}\,\Bigl ({\oq}{\sq}\,\Bigr )\,,\quad
W_{\Pi}=\ml {\rm Tr}\,\Pi_{\it ll}+\mh {\rm Tr}\,\Bigl (\Pi_{\it hl}\frac{1}{\Pi_{\it ll}}\Pi_{\it lh}\Bigr )\,,\label{(9.2)}
\eq
where $\hat\alpha(\mu)$ is the gauge coupling (with the pions $\Pi_{\it ll}$ sitting down inside), $\textsf{S}$ is the kinetic term of unhiggsed gluons,
$\Pi_{\it ll}$ is the $\nl\times\nl$ matrix of pions originated due to higgsing of ${\ov Q}_{\it l}, Q_{\it l}$ quarks, $\Pi_{\it lh}$ and $\Pi_{\it hl}$ are the $\nl\times\nh$ matrices of the hybrid pions (in essence, these are the quarks ${\ov Q}_{\it h}, Q_{\it h}$ with higgsed colors), $\oq$ and $\sq$ are the still active $\it h$-quarks with unhiggsed colors, and dots denote residual D-term interactions which are assumed to play no significant role in what follows and will be neglected.  $z^+_Q=z_Q(\la\,,\mgl)$ in \eqref{(9.2)} is the numerical value of the quark renormalization factor due to a perturbative evolution in the range of scales $\mgl<\mu<\la$,
\bbq
z^+_Q=z^+_Q(\la\,,\mgl)=z^+_Q(\langle\Pi^\dagger_{\it ll}\rangle,\langle\Pi_{\it ll}\rangle)\sim\Biggl (\frac{\mgl}{\la}\Biggr )^{\gamma_{+}}\ll 1\,,\quad \gamma_{+}=(2N_c-N_F)/(N_F-N_c)\,.
\eeq

The numbers of colors and flavors have already changed in the threshold region $\mgl/(\rm several)<\mu<(\rm several)
\mgl\,,\,N_F\ra {\hat N}_F=N_F-\nl=\nh \,,\,N_c\ra {\hat N}_c=N_c-\nl$\,, while the coupling $\hat \alpha(\mu)$ does not
change essentially and remains $\sim \alpha(\mu=\mgl)\gg 1$. Therefore, the new quark anomalous dimension $\gamma_{-}
(\hat N_c\,,\hat N_F=\nh\,,\hat a\gg 1)$ and the new $\hat\beta$ - function have the form
\bq
\frac{d {\hat a}(\mu)}{d\ln \mu}=-\,\, \nu_{-}\,\hat a(\mu)\,,\quad \nu_{-}=\frac{\hat N_F\, \gamma_{-}-{\hat b}_{o}}
{\hat N_c}=\frac{3\hat N_c-2\hat N_F}{\hat N_F-\hat N_c}=\nu_{+}-\frac{\nl}{N_F-N_c}\,, \label{(9.3)}
\eq
\bbq
{\hat b}_{o}=(3\hat N_c-\hat N_F)=(\bo-2\nl)\,,\quad  \gamma_{-}=\frac{2\hat N_c-\hat N_F}
{\hat N_F-\hat N_c}=\gamma_{+}-\frac{\nl}{N_F-N_c}\,\,.
\eeq

Depending on the value of $\hat N_F/\hat N_c$\,, the lower energy theory is in different
regimes. We consider only two cases below.\\

{\bf i)}\,\,\, {\boldmath $1<\hat N_F/\hat N_c<3/2$\,.}\quad In this case,\, $\nl<(3N_c-2N_F),\,\nu_{-}>0$,
and hence the coupling $\hat a(\mu)$ continues to increase with decreasing $\mu$\,, but more slowly than before.

The next physical scale is given by the pole mass of the active $\oq\,,\sq$  quarks
\bbq
\frac{\mhp}{\la}\sim\Biggl [\,\frac{\mh(\mu=\mgl)}{\la}\sim\frac{\mh}{\la}\Biggl (\frac{\la}{\mgl}
\Biggr )^{\gamma_{+}}\,\Biggr ]\Biggl (\frac{\mgl}{\mhp} \Biggr )^{\gamma_{-}}\quad \ra
\eeq
\bq
\frac{\mhp}{\mgl}\sim r\ll 1\,,\quad\quad \frac{\mhp}{\la}\sim\Bigl (\, r\,\Bigr )^{\nl/N_c}\Biggl (
\frac{\mh}{\la}\Biggr )^{(N_F-N_c)/N_c}\gg \frac{\lym}{\la}\,.\label{(9.4)}
\eq

After integrating out these $\oq\,,\sq$ quarks as heavy ones at $\mu<\mhp$, we are left with the $SU(\hat N_c)$
YM  theory  in the strong-coupling regime $a_{YM}(\mu=\mhp)\gg 1$ (and pions). The scale factor of its gauge coupling ${\hat\Lambda}_{YM}=\langle \Lambda_{L}(\Pi_{\it ll})\rangle$ is determined from
\bq
a_{YM}(\mu=\mhp)=\Biggl (\frac{\mhp}{{\hat\Lambda}_{YM}}\Biggr )^3={\hat a}(\mu=\mhp)=\Biggl (\frac{\la}{\mgl}
\Biggr )^{\nu_{+}}\Biggl (\frac{\mgl}{\mhp}\Biggr )^{\nu_{-}}\ra {\hat\Lambda}_{YM}=\lym. \label{(9.5)}
\eq

Finally, after integrating out the $SU(N_c-\nl)$ gluons at $\mu < \lym$ and if all pion fields, $\Pi_{\it hl}, {\Pi}_{\it lh}$ and $\Pi_{\it ll}$, do not evolve at $\mu<\mgl$, the Lagrangian of pions looks then at $\mu < \lym$ as
\bq
K=z^+_Q\Bigl (K_{\it ll}+K_{\rm hybr}\Bigr ),\,\, W=(N_c-\nl)\Biggl ( \frac{\la^{\bo}\mh^{\nh}}{\det \Pi_{\it ll}}\Biggr )^{\frac{1}{(N_c-\nl)}}+\ml {\rm Tr}\,\Pi_{\it ll}+\mh{\rm Tr}\Bigl (\Pi_{\it hl}\frac{1}{\Pi_{\it ll}}\Pi_{\it lh}\Bigr ).\label{(9.6)}
\eq

From this, the masses of $\Pi_{\it hl}\,,{\Pi}_{\it lh}$ and $\Pi_{\it ll}$ are
\bq
\mu(\Pi_{\it hl})=\mu({\Pi}_{\it lh})\sim \Bigl (r \Bigr )^{\gamma_{-}}\,\mhp\,,\quad
\mu(\Pi_{\it ll})\sim \frac{\ml}{z^+_Q}\sim r\, \mu(\Pi_{\it hl})\,.\label{(9.7)}
\eq

To check the self-consistency, i.e. that the active $\oq\,,\sq$  quarks are indeed in the HQ
phase and are not higgsed, we estimate the possible value of the gluon mass $\mgh$
\bq
\mgh^2\sim \Biggl [\,{\hat a}(\mu=\mgh)=\Biggl (\frac{\la}{\mgl} \Biggr )^{\nu_{+}}\Biggl
(\frac{\mgl}{\mgh}\Biggr )^{\nu_{-}}\,\Biggr]\langle {\oq}\sq\rangle_{\mu=\mgh}\,\,\,, \label{(9.8)}
\eq
\bq
\frac{\langle {\oq}\sq\rangle_{\mu=\mgh}}{\la^2}\sim\, r\,\frac{\mgl}{\la}\,\Biggl(\frac{\mgl}{\la}\Biggr )^{\gamma_{+}}
\Biggl ( \frac{\mgh}{\mgl}\Biggr )^{\gamma_{-}}\quad\ra \quad \mgh\sim \mhp\,\,\,, \label{(9.9)}
\eq
as could be expected. As before, we assume that $\mhp=(\rm several) \mgh$, see the footnote \ref{(f4)}.\\

{\bf ii)}\,\,\, {\boldmath $3/2<\hat N_F/\hat N_c<3$\,.}\quad In this case\, $\nu_{-}<0\,$\,,
and hence {\it the RG  flow is reversed} and the coupling $\hat a(\mu)$ starts to decrease
with decreasing $\mu$ at $\mu<\mgl$\,, approaching its fixed point value ${\hat a}_{*}<1$ {\it
from above} (unless it stops before at $\mu=\mhp$). Until ${\hat a}(\mu)\gg 1$, it behaves as
\bq
{\hat a}(\mu<\mgl)=a_{+}(\mu=\mgl)\Biggl(\frac{\mu}{\mgl} \Biggr)^{(-\nu_{-})>\,0}=\Biggl(\frac
{\la}{\mgl} \Biggr)^{\nu_{+}}\Biggl (\frac{\mgl}{\mu} \Biggr)^{\nu_{-}}\,, \quad \nu_{-}<0\,.\label{(9.10)}
\eq
It therefore decreases to $\sim 1$ at $\mu\sim \Lambda_o$,
\bq
{\hat a}(\Lambda_o)\sim 1\quad \ra \quad \frac{\Lambda_o}{\la}\sim \Biggl (\frac{\mgl}{\la}
\Biggr )^{\omega}\,,\quad \omega=\frac{\nl}{2{\hat N}_F-3{\hat N}_c}\,>1\,.\label{(9.11)}
\eq

We first consider the case $\mhp\gg\Lambda_o$. The value of $\mhp$ is then given by \eqref{(9.4)}, and this requires
\bq
\mhp\sim r\,\mgl \gg\Lambda_o\quad \ra \quad 1\gg r\gg r_3\,,\quad r_3\sim\Biggl (\frac{\mh}{\la}\Biggr )^{\frac{3N_c-2N_F}{2\nl}}\gg r_H\sim\Biggl (\frac{\mh}{\la}\Biggr )^{\frac{\nd}{N_c-\nl}}.\label{(9.12)}
\eq
Then the running of ${\hat a}(\mu)$ stops at ${\hat a}(\mhp)\gg 1$, and hence the theory does not
enter the conformal regime. The situation is here similar then to those described above in item
'{\bf i}'. The active quarks $\qh\,,\oqh$ decouple at $\mu<\mhp$, and the $SU({\hat N}_c)$
YM  theory remains (and pions) in the strong coupling regime, $a_{\rm YM}(\mu)=(\mu/\lym)
^3\gg 1$ at $\lym \ll \mu<\mhp$, etc. It can be dealt with as before in item '{\bf i}'.

The new regime is realized for the parameters values such that
$\mhp\ll \Lambda_o$, but still $\mgl\ll \la$. In this case, as the scale $\mu$ decreases below
$\mgl\ll \la$,  the large but decreasing coupling ${\hat a}(\mu\gg \Lambda_o)\gg 1$
crosses unity at $\mu=\Lambda_o$ and becomes ${\hat a}(\mu <\Lambda_o)< 1$, and the
theory enters the conformal regime, but with ${\hat a}(\mu)$ approaching its fixed point value
${\hat a}_*<1$ {\it from above}. The self-consistency of this regime then requires very
specific behavior of the quark anomalous dimension ${\hat \gamma}(\mu)=\gamma_{-}({\hat N}_F\,,
{\hat N}_c\,, {\hat a}(\mu))$ in the region $\mu\sim \Lambda_o$, when decreasing ${\hat a}(\mu)$
undershoots unity. Qualitatively, the behavior has to be as follows\,: a) ${\hat \gamma}(\mu)$
stays nearly intact at its value $(2{\hat N}_c-{\hat N}_F)/({\hat N}_F-{\hat N}_c)<{\hat b}_o
/{\hat N}_F$, as far as the coupling remains large, ${\hat a}(\mu\gg \Lambda_o)\gg 1$\,;\,\, b)
${\hat \gamma}(\mu)$ changes rapidly in the threshold region $\Lambda_o/(\rm several)<\mu<
(\rm several)\, \Lambda_o$. It begins to increase at $\mu=(\rm several)\,\Lambda_o$ and crosses
the value ${\hat b}_o/{\hat N}_F$ just at the point $\mu=\Lambda_o$, where ${\hat a}(\mu)$
crosses unity, such that the $\beta$ - function remains smooth and nonzero and does not change sign;
\,\, c) ${\hat \gamma}(\mu)$ continues to increase at $\mu<\Lambda_o$ and reaches its maximal
positive value at $\mu=\Lambda_o/(\rm several)$\,, and then begins to decrease with further
decreasing $\mu$\,, approaching its limit value (equal to ${\hat b}_o/{\hat N}_F$ at ${\hat b}_o>
0$\,, or zero at ${\hat b}_o<0$ \,) {\it from above} at $\mu\ll \Lambda_o$. It would be useful to
confirm this very specific behavior of ${\hat \gamma}(\mu)$ independently from elsewhere. But
once this is accepted, we can trace the lower energy behavior proceeding similarly to what
we did for the conformal regime (but additionally taking into account the presence of pions
which are remnants of {\it l} - quarks higgsed previously at the higher scale $\mgl$)\,.

\section{Conclusions}

{\hspace {0.5 cm} The mass spectra of ${\cal N}=1$ SQCD with $SU(N_c)$ colors and $N_F$
flavors of light quarks $Q\,,\ov Q$
(with masses $0<m_i\ll \la$) have been described above, within the dynamical scenario $\#2$.
This scenario implies that quarks can be in two different phases only\,:
the HQ (heavy quark) phase where they are confined, or the Higgs phase. Besides, we
have compared this (direct) theory with its Seiberg dual variant \cite{S1, IS}, which
contains $SU(N_F-N_c)$ dual colors, $N_F$ dual quarks $q\,,\ov q$\,  and $N_F^2$
additional mesons $M$ (mions).

As was shown above in the text, in those regions of the parameter space where an additional
small parameter is available (this is $0<\bo/N_F=(3N_c-N_F)/N_F\ll 1$ at the right end of the conformal window,
or its dual analog $0<\bd/N_F=(2N_F-3N_c)/N_F\ll 1$ at the left end), there are {\it parametrical differences}
in the mass spectra of direct and dual theories, and therefore they are clearly not equivalent. In fact, this implies
that even when both $\bo/N_F\sim \bd/N_F\sim 1$, there are no reasons for these two theories
to become exactly the same.
\footnote{\,
But to see the possible differences more clearly, it is insufficient in this case to make rough estimates of particle masses up to non-parametric factors $O(1)$ as has been done in this paper. One has either to resolve the mass spectra in more detail or to calculate some Green's functions in both theories and to compare them.
}

Besides, as was shown in section 5, one can trace unavoidable internal inconsistencies of the Seiberg duality in  variant "b", i.e. "confinement without chiral symmetry breaking" (this implies that at $N_c<N_F<3N_c/2$, the direct quarks and gluons form a large number of massive hadrons with masses $\sim \la$\,, while new light composite particles with masses $\mu_i\ll\la$ appear, described by the dual theory).
\footnote{\,
Similar problems with this variant "b" can also be traced in scenario $\#1$ \cite{ch1,ch2} (but in this
scenario, the differences between the direct and dual theories are much more pronounced), see also footnote \ref{(f7)}\,.
}
This agrees with some general arguments presented previously in section 7 in \cite{ch1} that the duality in the variant "b" cannot be realized ( it is also worth recalling that {\it those arguments were not related with the use of the scenario $\#1$ with the diquark condensate})\,.\\

As regards the mass spectra of the direct theory in the dynamical scenario of this paper, their main features are as follows (for $\bo/N_F=O(1)$\,, see Section 2 for $\bo/N_F\ll 1$).\\ 1) In all cases considered, there is a large number of gluonia with masses $\sim \lym=(\la^{\bo}\det m_Q)^{1/3N_c}$.\\
2) When all quark masses are equal, they are in the HQ  phase (i.e. not higgsed but confined, the string tension $\sqrt\sigma\sim\lym\lesssim m_Q^{\rm pole}$\,), for the whole interval $N_c<N_F<3N_c$\,, and hence form a large number of various hadrons with the mass scales\,: a) $\sim
m_Q^{\rm pole}\sim \lym$ at $3N_c/2<N_F<3N_c$\,, and \, b)\, $\sim m_Q^{\rm pole}\sim \la(m_Q/\la)^ {(N_F-N_c)/N_c}\gg \lym$ at $N_c<N_F<3N_c/2$. There are no additional lighter pions $\pi^{\ov j}_{i}$ with masses $\mu_{\pi}\ll m_Q^{\rm pole}$\,, for all $N_c<N_F<3N_c$\,.

3)  The case with $\nl$ flavors of smaller masses $\ml$ and $\nh=N_F-\nl$ flavors with larger masses $\mh\,,\,0<\ml<\mh\ll \la$ was also considered. When $\nl>N_c$\,, all quarks are also in the HQ  phase for all $N_c<N_F<3N_c$\,, and form a large number of hadrons whose masses
depend on their flavor content (see the main text), but there are no any additional lighter pions also.

4) Only when $\nl<N_c$\,, the {\it l} - flavored quarks $Q_{\it l}\,,{\ov Q}^{\it \ov l}$ are higgsed, $SU(N_c)\ra SU(N_c-\nl)$, and there $\nl^2$ lighter pions $\pi^{\it \ov l}_{\it l}$ appear, while the heavier {\it h} - flavored quarks $Q_{\it h}\,,{\ov Q}^{\it \ov h}$ always remain in the HQ  phase. In this case, the mass spectra and some new regimes with unusual properties of the RG  flow were presented in sections 7-9.\\

We have considered in this paper not all possible regimes, but only those that reveal some qualitatively new features. We hope that, if needed, a reader can deal with other regimes using the methods in \cite{ch1, ch2} and in this paper.\\

On the whole for this paper, the mass spectra of the direct theory with $SU(N_c)$ colors and $N_c<N_F<3N_c$ flavors of light quarks and its Seiberg's dual variant with $\nd=N_F-N_c$ dual colors have been obtained and compared in the framework of the dynamical scenario described in Introduction. The parametrical differences in the mass spectra of the direct and dual theories were traced for light quarks {\it with all nonzero masses}. Let us emphasize that, in all cases considered, no internal inconsistences were found in calculations of mass spectra both in the direct and dual theories. Let us recall also that the dynamical scenario used in this paper satisfies all those tests which were used as checks of the Seiberg hypothesis about the equivalence of the direct and dual theories. This shows, in particular, that all these tests, although necessary, may well be insufficient.\\

This work was supported in part by the Grant 14.740.11.0082 of the Federal Program "Personnel of Innovational Russia".

\appendix
\section{Appendix}

{\hspace {0.5 cm} The main purpose of this appendix is to show that the choice $\Lambda_q\sim \la$ used in the text is,
in essence, most favorable for the dual theory. But we first present a few useful formulas.

The RG-flow of the gauge coupling in the region $\mu_H \leq \mu \leq \la$, where $\mu_H\ll \la$ is the highest physical mass, is given by
\bq
\frac{da}{d\ln \mu}=\beta(a)=-\frac{N_F}{N_c}\,\frac{a^2}{1-a}\bigl (\Delta_o-\gamma_Q(a)  \bigr ),\quad
\ln\frac{\la}{\mu}=\frac{N_c}{N_F}\int_{a_{\Lambda}}^{a_\mu}\frac{da(1-a)}{a^2(\Delta_o-\gamma_Q)}\,,\label{(A.1)}
\eq
\bbq
0<\Delta_o=\frac{\bo}{N_F}\ll 1, \quad a_\mu=a(\mu)=\frac{N_c\alpha(\mu)}{2\pi},\quad a_*=\Delta_o+O(\Delta_o^2),
\eeq
\bbq
\quad a_{\Lambda}\equiv a(\mu=\la)=a_*(1-\delta),\quad 0<\delta\ll 1.
\eeq
Hence, using $\gamma_Q(a)\simeq a$, we obtain from \eqref{(A.1)} that for sufficiently small $\mu/\la\ll 1$,
\bq
a(\mu)= a_*\Bigl (1-\delta \epsilon_\mu\Bigr ),\quad \epsilon_\mu\approx\Bigl (\frac{\mu}{\la} \Bigr )^{3a^{2}_{*}}\approx
\Bigl (\frac{\mu}{\la} \Bigr )^{\frac{\rm b^2_o}{3N^2_c}}\,,\quad \mu_H<\mu\ll \la\,.\label{(A.2)}
\eq
Expression \eqref{(A.2)} can even be used as a reasonable interpolation in the whole interval $\mu_H<\mu<\la$. It follows from \eqref{(A.2)} that $a(\mu)$ approaches its fixed point value $a_*$ very slowly.

The RG-flow of the quark Kahler term renormalization factor from $z_Q(\la,s\mu=\la)=1$ down to $z_Q(\la,\mu\ll \la)\ll 1$
is given by
\bq
\gamma_Q(a)=\frac{d\ln z_Q}{d\ln \mu},\quad \gamma_Q(a)=\Delta_o-(a_*-a)+O\bigl ((a_*-a)^2  \bigr ).\label{(A.3)}
\eq
\bq
\ln\frac{1}{z_Q(\la,\mu)}=\frac{N_c}{N_F}\int_{a_{\Lambda}}^{a_\mu}\frac{da(1-a)}{a^2}\,\frac{\Bigl [\Bigl (\gamma_Q(a)-\Delta_o\Bigr )+
\Delta_o\Bigr ]}{(\Delta_o-\gamma_Q)}\,.\label{(A.4)}
\eq
\bq
z_Q(\la,\mu\ll \la)=\bigl (\frac{\mu}{\la}\bigr )^{\Delta_o}\rho\ll 1,\,\,\, \rho=\Bigl ({\frac{a_\Lambda}{a_\mu}}\Bigr )
^{\frac{N_c}{N_F}}\exp \Bigl \{\frac{N_c}{N_F}\Bigl (\frac{a_\mu-a_\Lambda}{a_\mu a_\Lambda }  \Bigr ) \Bigr \}\approx
\exp \Bigl \{\frac{\delta}{3\Delta_o}\Bigr \}\,.\label{(A.5)}
\eq

Clearly, the terms of the order $\sim \rho\sim\exp \{(\delta\ll 1)/\Delta_o\}$ are non-leading in comparison with the
terms $\sim \exp \{(c_o\sim 1)/\Delta_o \}\ll 1$, which are traced explicitly in the text (and we neglect such corrections in the main text).\\

We now consider the region $0<\Delta_o={\rm\bo}/N_F\ll 1$. Qualitatively, the value of the scale factor $\ld$ shows the characteristic scale where the logarithmic behavior of the dual gauge coupling ${\ov a}(\mu)={\ov N}_c{\ov \alpha}(\mu)/2\pi$  of the UV free dual theory at $\mu\gg \ld$ changes for the conformal freezing. At scales $\mu\sim \ld$, the coupling ${\ov a}(\mu\sim \ld)=O(1)$, the dual $\beta$ - function is also $O(1)$, and hence the dual theory enters quickly into the conformal regime (unlike the weakly coupled direct theory). We first consider the case $\ld\gg \la$ (the case $\ld\sim \la$ is considered in the main text). Then we can start to deal with the dual theory at the lower reference scale $\sim \la$, where it is already deep in the conformal regime, to match $\langle M(\mu=\la)\rangle=\langle{\ov Q}Q(\mu=\la)\rangle,
\,\,{\ov m}_Q(\mu=\la)=m_Q(\mu=\la)$ etc., and to proceed further exactly as in the text, and hence there will be no differences.

We now consider the case $\ld\ll \la$ (\,for example, $\ld/\la\sim\exp\{-1/(3\Delta^2_o)\}\,$,\, we recall that the number $\Delta_o\ll 1$, although small, {\it does not compete in any way} with the main small parameter $m_Q/\la$, see footnote 1), such that at $\mu\sim \la$, the dual theory is still deep in the logarithmic regime. In this case, in the interval of scales $\ld\ll\mu<\la$, the direct and dual theories are clearly different. By the definition of the scale $\la$, the direct theory entered already sufficiently deep into the conformal regime, i.e. $[a_*-a(\mu)]/a_*<\delta\ll 1$ at $\mu<\la$, while the dual theory is still deep in the logarithmic regime. We therefore consider lower energies $\mu\sim \ld$, where, by definition, even the dual theory entered sufficiently deep into the conformal regime. At this scale, some quantities of
the direct theory are given by
\bq
\widetilde{\cal M}_{\rm ch}^2=\langle{\ov Q}Q(\mu=\ld)\rangle\sim \Bigl (\frac{\ld}{\la}\Bigr )^{\frac{\bo}{N_F}}\Biggl ({\cal M}_{\rm ch}^2=\langle{\ov Q}Q(\mu=\la)\rangle\Biggr ),\label{(A.6)}
\eq
\bbq
\widetilde m_Q=m_Q(\mu=\ld)\sim \Bigl (\frac{\la}{\ld}\Bigr )^{\frac{\bo}{N_F}}m_Q(\mu=\la),\quad
m_Q(\mu=\la){\cal M}_{\rm ch}^2=\widetilde m_Q\widetilde{\cal M}_{\rm ch}^2=\lym^3\,,
\eeq
and therefore \eqref{(3.1)} at the scale $\mu\sim \ld\ll \la$ has the same form, with the normalizations: $M(\mu=\ld)=\widetilde{\cal M}_{\rm ch}^2$ and ${\ov m}_Q(\mu=\ld)=\widetilde m_Q$, while $\mu_1$ in \eqref{(3.1)} is now rewritten as $\mu_1\equiv {\widetilde Z}_q \ld$. After this, all calculations are the same as in section 3 with only notational changes. The only point that deserves additional comment is the explicit form of $\lym$, see \eqref{(2.2)},\eqref{(2.1)}, which was used in \eqref{(3.3)},\eqref{(3.4)} for finding $Z_q$. But it can be seen from \eqref{(A.6)} that $\lym$ stays intact.
The same can be seen from the expression in \cite{sv2} (see also the review \cite{SV-r})
for the gluino condensate of the direct theory in terms of the running scale $\mu$\,
\bq
\langle S\rangle=\mu^{\bo/N_c}m_Q^{N_F/N_c}(\mu)\frac{1}{a(\mu)}\exp\{-\frac{1}{a(\mu)}\},
\quad \langle S\rangle =\frac{\langle\lambda\lambda\rangle}{32\pi^2}=\lym^3\,,\label{(A.7)}
\eq
which is valid from sufficiently large $\mu$ down to $\mu=\mu_H\ll \la$, where $\mu_H$ is the largest physical mass. Now, taking $\mu=\la$ in \eqref{(A.7)}, we can write (with our exponential accuracy and neglecting $\delta\ll 1$ in comparison with unity in $a(\mu=\la)=a_*(1-\delta)\approx \Delta_o\approx{\rm \bo}/3N_c$\,):
\bq
\lym^3\sim \la^{\frac{\bo}{N_c}}\Bigl ( m_Q=m_Q(\mu=\la)\Bigr )^{\frac{N_F}{N_c}}\Biggl [\exp\{-\frac{1}{a(\mu=\la)}\}\sim
\exp\{-\frac{3N_c}{\rm\bo} \} \Biggr ]. \label{(A.8)}
\eq
On the other hand, taking $\mu=\ld$ in \eqref{(A.7)}, we obtain
\bq
\lym^3\sim \ld^{\frac{\rm\bo}{N_c}}\Bigl (\widetilde m_Q=m_Q(\mu=\ld) \Bigr )^{\frac{N_F}{N_c}}\Biggl [ \exp\{-\frac{1}{a(\mu=\ld)}\}\sim \exp\{-\frac{3N_c}{\rm\bo} \} \Biggr ].\label{(A.9)}
\eq
Hence, instead of $\la$ and $m_Q=m_Q(\mu=\la)$ in \eqref{(A.8)}, \,$\lym$ can equivalently be expressed through
$\ld$ and $\widetilde m_Q=m_Q(\mu=\ld)$ in \eqref{(A.9)}.

Therefore, we obtain the same result: ${\widetilde Z}_q=Z_q\sim\exp\{-N_c/\bo \}$, and at $\mu<\ld$ all results for observable masses in Section 3 will stay intact (with only notational changes). Nevertheless, in a sense, this variant with $\ld\ll \la$ is worse for the dual theory in comparison with $\ld\sim \la$ because both theories are additionally not equivalent in the interval of energies $\ld<\mu<\la$.

We now consider the region $0<{\rm\bd}/N_F\ll 1$ where weakly coupled is the dual theory. Here, we can simply repeat all the above arguments to see that all the results in section 4 remain valid, except for a change of notation as in \eqref{(A.6)} and the omission of the exponential factors in \eqref{(A.8)},\eqref{(A.9)} in accordance with the direct theory being strongly coupled here, $a_*=O(1)$. The same applies to the results for observable masses in Section 5, because all these can be expressed in terms of $\lym$ and the appropriate powers of $r=\ml/\mh$ and ${\widetilde Z}_q=Z_q\sim\exp\{-\nd/7{\rm\bd} \}$.

\end{document}